\documentclass[11pt, letterpaper]{article}

\usepackage{amsmath,amssymb,mathtools}
\usepackage{graphicx}
\usepackage{booktabs}
\usepackage{lineno}
\usepackage{algorithm}
\usepackage{algorithmic}
\usepackage{caption}
\usepackage{float}
\usepackage{subcaption}
\usepackage{siunitx}
\usepackage{microtype}
\usepackage{tabularx}
\usepackage{tikz}
\usetikzlibrary{calc, positioning}
\usetikzlibrary{arrows.meta}
\graphicspath{{Figures/}}
\usepackage[colorlinks=true,linkcolor=blue,citecolor=brown,urlcolor=brown]{hyperref}

\usepackage[margin=1in]{geometry}
\usepackage{authblk}
\usepackage[numbers,sort&compress]{natbib}
\title{Probabilistic Hazard Analysis Framework with Stochastic Optimal Control for Deteriorating Civil Infrastructure Systems}

\author[1]{Sudhir P. Jodha}
\author[1]{Konstantinos G. Papakonstantinou\thanks{Corresponding author: kpapakon@psu.edu}}
\affil[1]{Department of Civil and Environmental Engineering, The Pennsylvania State University, University Park, PA 16802, USA}

\date{}

\begin{document}

\maketitle

\begin{abstract}
\noindent The safety and resilience of civil infrastructure systems are increasingly threatened by compounded risks from various hazard events and structural deterioration due to environmental stressors. This study presents a comprehensive risk-informed, life-cycle optimization framework that extends the Performance-Based Earthquake Engineering (PBEE) and probabilistic seismic loss estimation paradigms by combining hazard uncertainties, nonstationary deterioration, structural damage accumulation, and state-dependent fragility assessments, with optimal, adaptive maintenance strategies in time. The life-cycle cost optimization is formulated in this work as a Markov Decision Process (MDP) problem, utilizing derived transition matrices reflecting time-variant deterioration effects and hazard risks. To mitigate the curse of dimensionality in system-level optimization, a novel tensor-based method exploiting Kronecker-factored transition dynamics is introduced, reducing complexity from exponential to linear in the number of components for separable transition dynamics, while still preserving exact, global dynamic programming solutions. The same formulation is shown to also be able to accommodate spatially correlated hazards. Overall, the framework is general and versatile, able to accommodate various hazard types. A seismic hazard application is, however, demonstrated and explained in detail in this work. The developed methodology eventually provides decision-makers with a practical, data-driven tool toward cost effective risk mitigation of civil infrastructure systems.
\end{abstract}

\vspace{1em}
\noindent \textbf{Keywords:} Hazard analysis, Nonstationary deterioration, Markov decision process, Dynamic Bayesian network, State-dependent generalized fragility, Tensor-based value iteration, Performance-based earthquake engineering

\section{Introduction}
\subsection{Overview}
\noindent
Traditional infrastructure hazard analysis frameworks are essential for managing complex systems and ensuring their safety and operational continuity. However, their inherently static nature is often insufficient for capturing time-dependent degradation, evolving environmental hazards, and the natural evolution of the systems driven by structural interventions in time. This paper addresses this critical gap by transforming the classical risk and hazard analysis formulation into a dynamic one. Crucially, our framework effectively captures the physics of nonstationary deterioration and damage accumulation, while also uniquely integrating them into a stochastic optimal control methodology. Such an approach is timely, given the accelerating degradation of infrastructure systems \citep{asce2025reportcard,infrastructure2021comprehensive,asce2021failure} and ongoing environmental shifts \citep{milly2008stationarity, bender2022corrosion, wang2012impact, stewart2011climate, nguyen2013assessment, sousa2020expected,bender2010modeled}. While comprehensive decision-making frameworks have been suggested \citep{biondini2006probabilistic, barone2014optimization, de2022climate, frangopol2024lifecycle}, managing this complex evolution requires two critical capabilities: a rigorous method to quantify the evolving risk and a systematic approach to optimally control it through interventions. In this work, we thus introduce a time-aware, probabilistic framework to globally optimize the life-cycle cost of infrastructure systems under hazards, within a risk-informed paradigm.

\begin{figure*}[t]
  \centering
  \includegraphics[width=\textwidth]{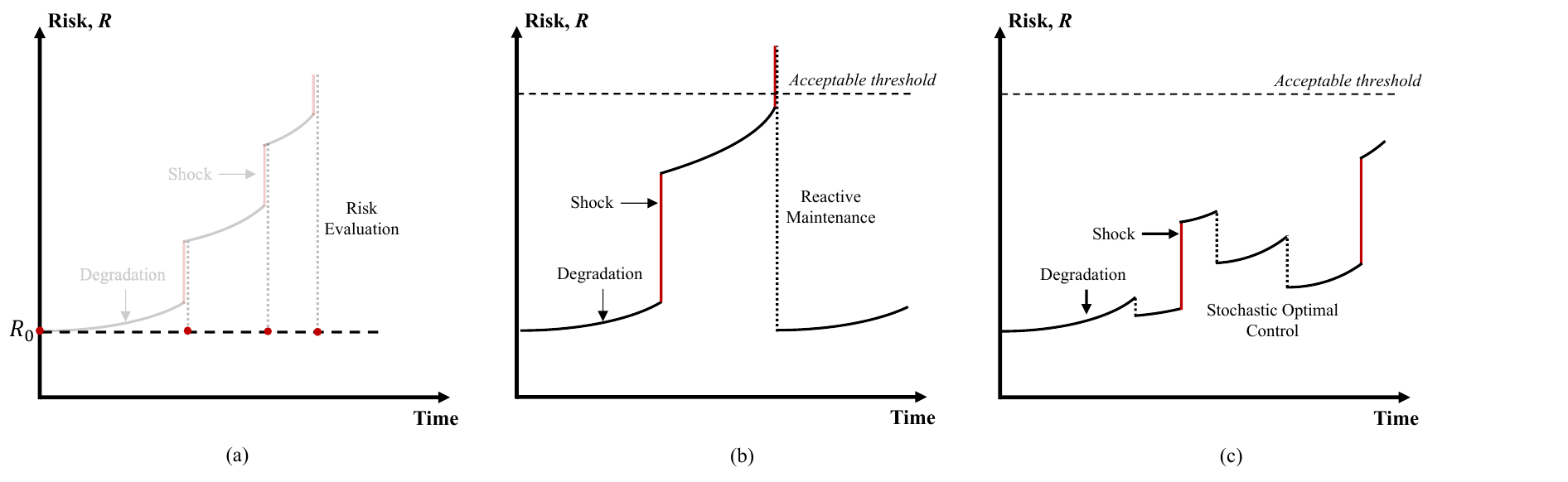}
  \caption{Conceptual illustration of risk evolution under three hazard analysis and maintenance strategies: 
  (a) Static Risk Assessment, illustrating conventional PEER PBEE methodology, where discrete evaluations (red dots) assume the system remains at a constant pristine baseline risk, disregarding continuous background accumulation of deterioration and shock damage; 
  (b) Reactive Maintenance, where interventions are triggered only when risk exceeds an acceptable threshold due to progressive deterioration and sudden shocks; and 
  (c) the proposed Stochastic Optimal Control framework, which explicitly accounts for time-dependent deterioration and derives state-dependent, preventive interventions to optimally manage risk over the life-cycle.}
  \label{fig:performance_diagrams}
\end{figure*}
Although the proposed framework is mathematically general and hazard-agnostic, the seismic domain provides an ideal foundational context due to its mature risk assessment methodologies. Within this domain, a key development in establishing quantitative risk objectives is Performance-Based Earthquake Engineering (PBEE), which has become the modern approach for seismic assessment by shifting the focus from prescriptive code compliance to achieving explicit performance objectives \citep{moehle2004framework, krawinkler2005van}. While early PBEE methodologies established a crucial link between design and performance, they were often based on deterministic criteria or qualitative performance levels. The framework developed by the Pacific Earthquake Engineering Research (PEER) Center represents a significant methodological advancement by introducing a fully probabilistic approach \citep{moehle2004framework, porter2003overview}. This PEER PBEE methodology rigorously propagates uncertainties to quantify performance in terms meaningful to stakeholders, such as economic loss and downtime \citep{gunay2013peer, yang2009seismic, gardoni2002probabilistic, castillo2022functionality,asce2024effects}. To achieve this, the canonical framework decomposes the assessment into four sequential stages linked by the theorem of total probability:
\begin{equation}
\begin{aligned}
\lambda(DV) 
&= \int_{DM}\!\!\int_{EDP}\!\!\int_{IM}
  \underbrace{f(DV\mid dm)}_{\text{Loss}}\,
  \underbrace{f(dm\mid edp)}_{\text{Damage}} 
  \underbrace{f(edp\mid im)}_{\text{Structural}}\,
  \underbrace{f(im)}_{\text{Hazard}}\,
  \mathrm{d}dm\,\mathrm{d}edp\,\mathrm{d}im ,
\end{aligned}
\label{eq:peer}
\end{equation}
where $\lambda(DV)$ is the mean annual frequency of exceeding a Decision Variable ($DV$) threshold (e.g., economic loss or downtime). In Eq.~\eqref{eq:peer}, $f(\cdot)$ generally represents relevant probability density functions (PDF). $\lambda(DV)$ is computed by integrating the conditional probabilities of the $DV$ given a Damage Measure ($DM$, e.g., component failure), the $DM$ given an Engineering Demand Parameter ($EDP$, e.g., inter-story drift), the $EDP$ given a hazard Intensity Measure ($IM$, e.g., peak ground acceleration), and the annual frequency of the hazard intensity $f(im)$ \citep{sullivan2014simplified, moehle2004framework}. This modular structure allows for a rigorous, transparent propagation of uncertainty throughout the analysis.

However, the PEER PBEE framework is still conventionally static in nature, evaluating structures in a single condition state, often by assuming a pristine, as-built condition when determining $f(edp \mid im)$ \citep{krawinkler20049, porter2003overview,ramirez2012expected}. Consequently, it provides a static snapshot of risk rather than a dynamic trajectory, and does not explicitly account for continuous degradation, damage accumulation, or the historical trajectory of the asset as it is continuously reshaped by interventions over its extended lifespan.

To understand the implications of this static assumption, Figure~\ref{fig:performance_diagrams} conceptualizes the evolution of risk, driven by two types of hazard processes: (1) progressive degradation, capturing time-dependent deterioration from phenomena such as corrosion, and (2) shock-based hazard events, representing sudden increases in risk \citep{mori1994maintaining, sanchez2011life, ghosh2010aging, frangopol2011life}. Figure~\ref{fig:performance_diagrams}a, representing a ``Static Risk Assessment," approach illustrates the traditional PEER PBEE perspective. In this idealized representation, $R_0$ denotes the baseline risk level assessed here at the initial, as-built condition. Importantly, this diagram highlights the limitation of the traditional framework, which only evaluates the structure always assuming this initial condition state. Consequently, subsequent risk evaluations over time (indicated by discrete assessment points, shown as red dots) effectively assume the vulnerability remains constant and identical to $R_0$. This disregards the true, continuous accumulation of background degradation and shock damage (shown by the faded gray curve) and/or always assumes perfect repairs/replacements in time, failing to capture the true evolution of performance and, by remaining a static snapshot, offering no mechanism to optimally control risk over time.

Meanwhile, the maintenance and life-cycle engineering literature has explicitly addressed this time dimension. Numerous studies have focused on condition-based and reactive strategies to manage structural aging \citep{sanchez2011life,sanchez2016maintenance,biondini2016life, barone2014optimization}. Figure~\ref{fig:performance_diagrams}b depicts a standard reactive maintenance approach, where interventions are triggered once the risk exceeds a determined threshold. While this strategy effectively limits peak vulnerability, it inherently relies on heuristic rules for determining appropriate, relevant thresholds. More importantly, these time-dependent maintenance models typically fail to embed the hazard in a fully probabilistic, risk-based form, such as in a rigorous Probabilistic Seismic Hazard Analysis (PSHA) \citep{baker2021seismic}.

This highlights a fundamental disconnect in the current literature: the PEER PBEE framework lacks a mathematical mechanism for time effects and interventions, while time-dependent maintenance models and frameworks fail to embed hazards in a practical, probabilistic manner, following established approaches, like the PEER PBEE one. To bridge this gap, this work uniquely extends the PEER PBEE framework into a dynamic environment that integrates hazard uncertainties, nonstationary deterioration, and structural damage accumulation, all within a stochastic optimal control framework. As illustrated in Figure~\ref{fig:performance_diagrams}c, the proposed framework ensures that interventions are optimally timed and sized based on the evolving risk state. Rather than relying on discrete, static baseline assumptions (Figure~\ref{fig:performance_diagrams}a) or reacting to heuristic threshold exceedances (Figure~\ref{fig:performance_diagrams}b), the resulting maintenance policy is state-dependent, hazard-aware, and globally optimal, dynamically balancing system reliability with life-cycle cost.

Over the past decades, researchers have addressed various pieces of this complex problem. Significant progress has been made in characterizing stochastic hazard risks, particularly through the maturation of probabilistic hazard analysis and regional loss estimation methods \citep{mcguire2004seismic, baker2021seismic, coburn2002earthquake, heresi2023rpbee, laadaptation}. Parallel to these hazard-centric advancements, a rich body of literature exists on probabilistic modeling of structural deterioration and life-cycle reliability \citep{anwar2024life}. Numerous studies have modeled the progressive degradation of infrastructure components due to mechanisms like corrosion, fatigue, and environmental attacks, often treating deterioration as a stochastic process \citep{mori1994maintaining,ellingwood1997reliability,jia2018stochastic, sanchez2011life, van2009survey, papakonstantinou2013probabilistic,dehghani2021markovian, lin2019integrative}. Concurrently, researchers have also developed time-dependent fragility analyses to quantify how structural vulnerability to hazards changes in time \citep{ghosh2010aging, SAYDAM2013221}. For example, Choe et al. \citep{choe2009seismic} showed that corrosion-induced section loss can substantially elevate the seismic fragility of reinforced concrete bridges over time. Esteva et al. \citep{esteva2016structural} likewise investigated life-cycle seismic performance of buildings, demonstrating the importance of accounting for damage accumulation across multiple earthquakes. Similarly, Iervolino et al. \citep{iervolino2016markovian} developed a stochastic framework to model this damage accumulation as a Markovian process. Andriotis and Papakonstantinou \citep{andriotis2018extended} introduced generalized fragility formulations using a dependent Markov model to generally handle time-dependent effects in structural vulnerability analysis. Along these lines, recent efforts have also leveraged Dynamic Bayesian Networks (DBN) to efficiently derive state-dependent seismic fragility functions \citep{nardin2025uq}. Notably, Molaioni et al. \citep{molaioni2024dynamic, MOLAIONI2026102654} have developed a detailed DBN to evaluate life-cycle fragility for corroding bridges. Broadening this scope, Otárola et al. \citep{otarola2024multi} introduced a comprehensive Markovian framework for multi-hazard life-cycle consequence analysis. While their approach effectively captures the dynamic evolution of risk and accumulating damage, it remains focused on consequence assessment rather than the explicit integration with optimal control policies.

Another important relevant research direction has focused on optimal maintenance strategies under uncertainty. This includes building on life-cycle cost optimization frameworks and adaptive planning heuristics \citep{sabatino2015sustainability, madanat2006adaptive}. To systematically formalize this sequential decision-making process, optimal inspection and maintenance planning has been widely treated with Markov Decision Processes (MDP) and solved via dynamic programming \citep{papakonstantinou2014planning, morato2022optimal}. In this context, DBNs have proven effective for modeling complex system-level dependencies and supporting adaptive, risk-based inspection and maintenance strategies \citep{luque2019risk, bismut2021optimal}. To further address the curse of dimensionality inherent in large-scale systems, recent approaches have leveraged Deep Reinforcement Learning (DRL) to find near-optimal policies in vast state and action spaces \citep{andriotis2019managing,andriotis2021deep, fan2023deep}. Notably, Saifullah et al. \citep{saifullah2022,saifullah2026multi} demonstrated the potential of multi-agent DRL to guide maintenance decisions for large networks of deteriorating bridges and pavements, while Morato et al. \citep{morato2023inference} combined DBN deterioration modeling with multi‑agent reinforcement learning to derive system‑level inspection and maintenance policies. Building on this direction, Metwally et al. \citep{metwally2024managing} frame the problem of managing aging bridges under seismic hazards as a Partially Observable Markov Decision Process (POMDP), solved by multi-agent DRL, and they utilize a DBN to explicitly model the evolution of state-dependent seismic fragility, a concept also central to our framework here. 

All these contributions provide valuable building blocks for a comprehensive probabilistic hazard analysis and life-cycle management framework. Yet, a unified time-dependent hazard analysis framework incorporating life-cycle interventions and able to provide globally optimal solutions is still lacking. This work bridges this gap by recasting the established probabilistic hazard analysis (PHA) paradigm, such as PSHA in the seismic setting, from a static risk quantification tool into a dynamic, decision-theoretic framework that jointly captures the evolving risk profile and determines globally optimal interventions over the system's lifespan. To realize this, the hazard dynamics and time-dependent deterioration are modeled within a DBN. This probabilistic graphical model explicitly captures the system's temporal evolution, and can support state-dependent generalized fragility functions \citep{andriotis2018extended} that quantify the variation in structural vulnerability due to aging and accumulated damage. Building upon this probabilistic foundation, the life-cycle management problem is then formally cast as a MDP. Within this formulation, the system states describe the structure's evolving condition driven by the DBN-derived transition probabilities, while the available actions represent structural interventions. Notably, the inherent Markov assumption is not a limitation here, as the state space can be strategically augmented to incorporate historical dependencies, as needed \citep{papakonstantinou2014planning, morato2022optimal, morato2023inference}. The resulting architecture is also mathematically versatile and can easily accommodate various hazard types.

A significant challenge in this framework is the curse of dimensionality in solving the devised MDP, particularly when the infrastructure system is modeled as a portfolio of multiple structures. The combinatorial growth of the state space can render standard dynamic programming methods computationally intractable. To mitigate this while still preserving global optimality, we introduce a novel tensor-based solution algorithm that utilizes tensor algebra and the Kronecker product to formally exploit the problem's inherent factored structure. This approach enables higher scalability, while preserving the exact dynamic programming solution, and reduces the computational complexity from exponential to linear in the number of system components for separable transition dynamics, and can even extend to partially coupled dynamics, often induced by spatially correlated hazards.

Overall, the presented approach in this work provides detailed, optimal solutions that balance time-dependent effects, hazard risks, and maintenance costs for infrastructure systems, substantially advancing current probabilistic hazard analysis and performance-based engineering approaches. The remainder of this paper is structured as follows: Section 2 details the methodology and mathematical formulations. Section 3 formulates the life-cycle hazard analysis optimization problem as a MDP and presents the tensor-based solution algorithm. Section 4 demonstrates the framework through a numerical example, followed by a discussion of the results and conclusions in Section 5.

\section{Methodology}\label{sec:methodology}
\noindent
The overarching objective of the proposed framework is to mathematically formalize the life-cycle optimization of infrastructure systems, explicitly integrating time-dependent hazard risks, progressive degradation, and system-level interactions. The methodology is formulated through five interconnected modules: (1) a probabilistic hazard analysis to characterize the frequency and intensity of hazards; (2) a nonstationary deterioration model capturing the continuous, time-dependent physical degradation of the system; (3) a structural analysis workflow simulating the asset's response to discrete hazard events; (4) a damage accumulation model quantifying the impact of these discrete events on the structural condition; and (5) a state-dependent fragility analysis evaluating the evolving vulnerability over the asset's lifespan. Together, the integration of these components provides the rigorous physical and probabilistic parameters necessary to formulate the dynamic risk assessment and stochastic optimal control methodology detailed in subsequent sections.

\subsection{Probabilistic Hazard Analysis}
\label{sec:hazard}
\noindent
As mentioned previously, the framework is designed to be hazard-agnostic. Without loss of generality, while the demonstration focus is on seismic hazard, the methodology can accommodate any hazard type and $IM$ definition, as well as multi-dimensional vectors where several $IM$s can be combined to better characterize the hazard demand. For this study, we concentrate on seismic hazards due to their extensive characterization in the existing literature and their significant risk to civil infrastructure. The seismic hazard at the structure's location is defined by a hazard curve, which characterizes the annual frequency of exceeding various levels of ground motion intensity. This curve is the outcome of a PSHA, which integrates multiple components like the regional seismicity (e.g., mapped faults and source zones), site-specific conditions (such as soil type and amplification effects), and ground motion prediction equations (GMPEs) that quantify how seismic energy attenuates with distance and local geology. Together, these elements yield a probabilistic representation of ground shaking intensity at a specific location \citep{baker2008introduction, baker2021seismic}.

\begin{figure}[tb]
    \centering
    \includegraphics[width=0.45\columnwidth]{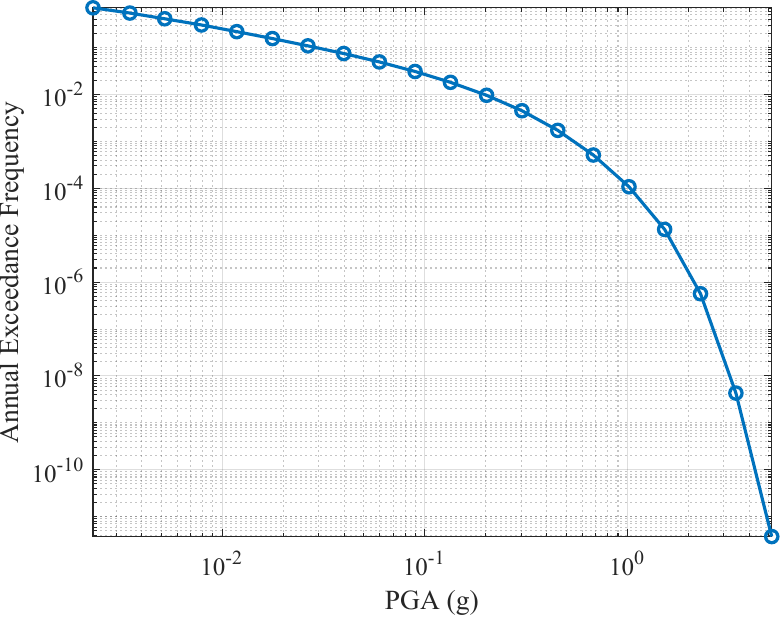}
    \caption{Seismic hazard curve for a specific site, based on data from the U.S. Geological Survey (USGS) \citep{usgs2025hazard}.}
    \label{fig:hazard_curve}
\end{figure}

The United States Geological Survey (USGS) maintains a comprehensive seismic hazard database that encapsulates this process, producing location-specific hazard curves \citep{usgs2025hazard}. These curves are publicly available and widely accepted in both academic and engineering practice and are utilized in our case study. A representative example is shown in Figure~\ref{fig:hazard_curve}. This curve plays a key role in the overall risk assessment process, serving as the probabilistic basis for sampling ground motion intensities over the structure’s life-cycle. Within the PEER PBEE loss estimation framework expressed in Equation~\ref{eq:peer}, the hazard curve relates to the PDF, $f_{IM}(im)$, which describes the likelihood of different IM occurring and thus directly influences the upstream structural response, damage, and loss estimations \citep{Clayton2023_NSHMapps, sullivan2014simplified, moehle2004framework}.

For spatially distributed infrastructure systems, such as transportation networks or building portfolios, the assumption of independent ground motions at different locations during a single earthquake event is not accurate. The correlation of ground motion intensities between sites is known to be strong for nearby locations and decays with increasing separation distance \citep{goda2008spatial}. To model this critical phenomenon, we employ the spatial correlation model developed by Jayaram and Baker (2009) \citep{jayaram2009correlation}. This model specifically addresses the correlation of intra-event residuals, which is the random component of ground motion variability not captured by standard GMPEs. The model provides an empirically derived formula for the correlation coefficient, $\rho(d)$, between the intra-event residuals of logarithmic $IMs$ at two sites separated by a distance $d$:
\begin{equation}
\rho(d) = \exp\left(-\frac{3d}{r}\right)
\label{eq:jayaram_baker}
\end{equation}
where $r$ is a decay parameter, often referred to as the correlation range, which depends on factors such as the spectral period of the $IM$ and local site conditions ($V_{s30}$). 

To simulate a seismic event affecting the system, a spatially correlated field of $IM$ values can be generated for all $n$ locations using a procedure based on the Nataf transformation \citep{derkiureghian1986,vlachos2018predictive}. First, a $n \times n$ correlation matrix $\mathbf{R}$ is constructed, where each element $R_{ij}$ is calculated using Equation~\ref{eq:jayaram_baker}. Next, this matrix is decomposed using Cholesky decomposition to find a lower triangular matrix $\mathbf{L}$ such that $\mathbf{R} = \mathbf{L}\mathbf{L}^T$. A vector $\mathbf{z}$ of $n$ independent standard normal random variables is then sampled and transformed into a vector of correlated standard normal variables, $\mathbf{y}$, via the linear transformation $\mathbf{y} = \mathbf{L}\mathbf{z}$. Finally, each component $y_k$ of this vector is transformed from the standard normal space back into the original $IM$ space using the inverse of the site-specific CDF, $IM_k = F_{IM_k}^{-1}(\Phi(y_k))$, where $\Phi(\cdot)$ is the standard normal CDF. This ensures the generated $IM$ values are consistent with local hazard curves while preserving the required spatial correlation.

\subsection{Deterioration Model}\label{sec:det}
\subsubsection{Overview}
\noindent
We model the gradual accumulation of structural damage due to environmental factors, using corrosion as a representative and critical deterioration mechanism. The corrosion-induced section loss, denoted as $D(\tau)$ at time $\tau$, is modeled as a nonstationary gamma process. It is a continuous-time, continuous-state stochastic process characterized by independent, non-negative increments, making it well-suited for modeling monotonic deterioration phenomena such as corrosion \citep{papakonstantinou2013probabilistic, van2009survey}. For integration into the MDP framework, the process is evaluated at discrete time intervals $\Delta\tau$. Unlike a stationary process where the rate of deterioration is constant, a nonstationary process allows the deterioration rate to change over time (e.g., accelerate), which accurately represents many corrosion mechanisms where protective layers break down or environmental aggressiveness changes \citep{mahmoodian2014modeling, nettis2024corrosion}.

\begin{figure}[t]
    \centering
    \includegraphics[width=0.45\columnwidth] {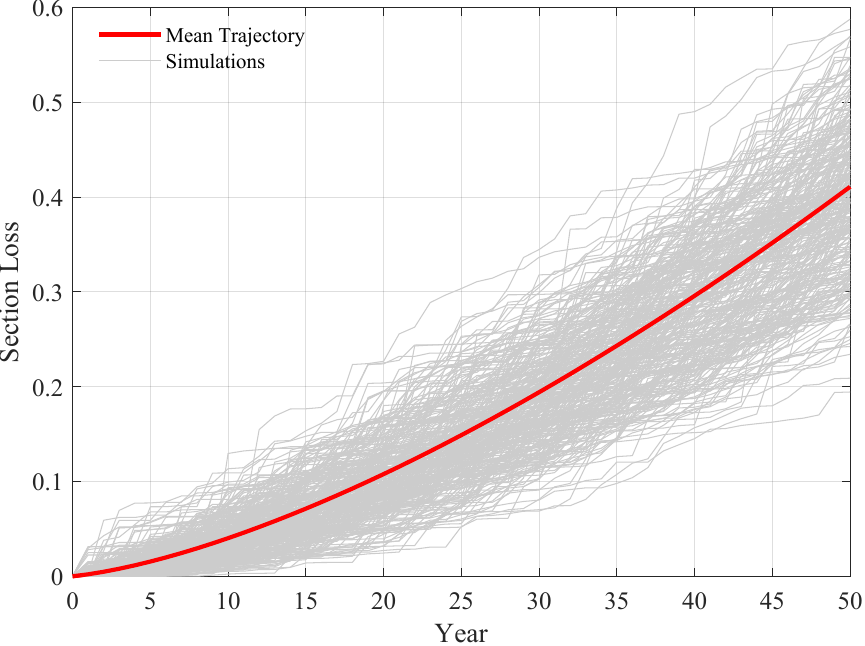}
    \caption{Simulated trajectories of the nonstationary gamma process for section loss due to corrosion. The solid red line indicates the mean deterioration path, $\mathbb{E}[D(\tau)]$, while the gray lines represent individual stochastic realizations.}
    \label{fig:gamma_process_traj}
\end{figure}

Mathematically, a stochastic process $\{D(\tau), \tau \ge 0\}$ is a nonstationary gamma process if it starts at zero ($D(0)=0$), has independent increments, and the increment over any time interval $(\tau_1, \tau_2]$, $\Delta D = D(\tau_2) - D(\tau_1)$, follows a gamma distribution. This distribution's shape parameter is given by the change in a shape function, $\alpha(\tau_2) - \alpha(\tau_1)$, while its rate parameter $\beta$ is constant. Consequently, the distribution of the total accumulated deterioration at time $\tau$ is given by a gamma distribution, $D(\tau) \sim \text{Gamma}(\alpha(\tau), \beta)$. The PDF is:
\begin{equation}
    f_{D(\tau)}(x) = \frac{\beta^{\alpha(\tau)}}{\Gamma(\alpha(\tau))} x^{\alpha(\tau)-1} e^{-\beta x}, \quad x \ge 0
    \label{eq:gamma_pdf}
\end{equation}
where $\Gamma(\cdot)$ is the gamma function. The mean and variance of the distribution are $\mathbb{E}[D(\tau)] = \alpha(\tau)/\beta$ and $\text{Var}[D(\tau)] = \alpha(\tau)/\beta^2$, respectively. To capture the nonstationary behavior, the shape function $\alpha(\tau)$ is defined as a non-linear function of time. A common and flexible choice is the power-law form \citep{van2009survey}:
\begin{equation}
    \alpha(\tau) = a \tau^b
    \label{eq:power_law}
\end{equation}
\textcolor{black}{where parameters $a$ and $b$ are case-specific and depend on the environmental exposure and the structural material properties. They can be calibrated from
physical corrosion models or empirical data. The power-law form adopted here is only representative; other deterioration models can be used without any modification to the subsequent formulation and no loss of generality.} Figure~\ref{fig:gamma_process_traj} illustrates typical realizations of this nonstationary process with solid red line indicating the mean deterioration path. 

\subsubsection{Deterioration Transition Formulation}\label{sec:det_form}
\noindent
For integration into the MDP framework, the continuous deterioration level $D(\tau)$ must be discretized into a finite number of $n_{CDS}$ mutually exclusive and collectively exhaustive Corrosion Damage States (CDS). These states are defined by $n_{CDS}-1$ thresholds, where CDS $l$ corresponds to the condition $c_{l-1} \le D(\tau) < c_l$, with boundary conditions $c_0=0$ and $c_{n_{CDS}}=\infty$.

\begin{figure*}[t]
    \centering
    \includegraphics[width=0.7\textwidth]{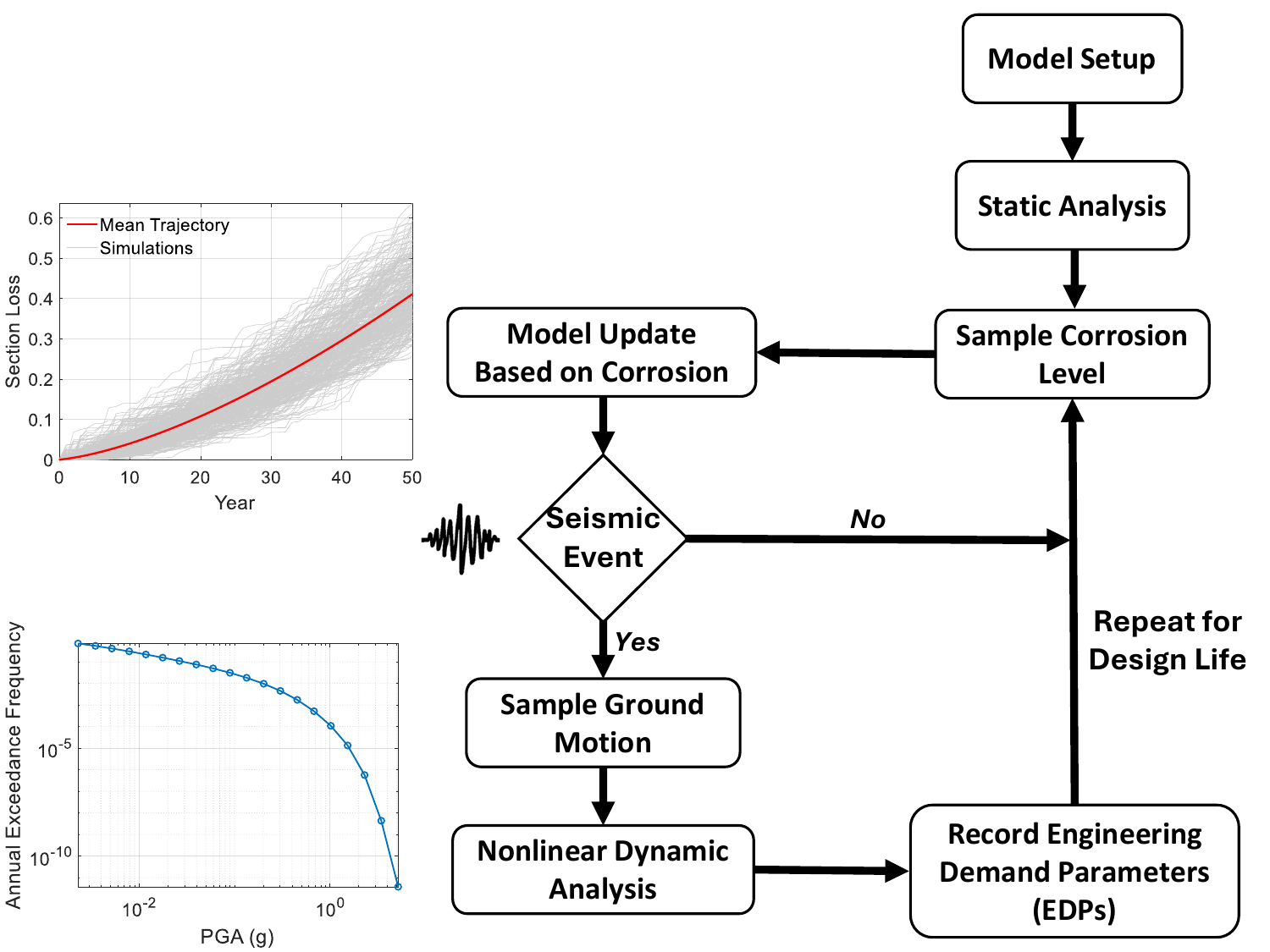}
    \caption{Flowchart of the structural simulation workflow for comprehensive performance assessment.}
    \label{fig:workflow}
\end{figure*}

A critical challenge that arises in this discretization is that the resulting discrete-state process is not inherently Markovian. The probability of transitioning from state $l$ to state $m$ depends on the precise damage level $D(\tau)$ within the interval $[c_{l-1}, c_l)$, not just on being in state $l$. To derive the exact one-step transition probability, $P_{l \to m}(\tau, \Delta\tau)$, we must therefore average the transition probabilities over all possible starting damage levels within state $l$, weighted by their likelihood. This is achieved rigorously using the law of total probability.

The exact transition probability is given by the integral:
\begin{equation}
\begin{aligned}
    P_{l \to m}(\tau, \Delta\tau) &= \int_{c_{l-1}}^{c_l} f_{D(\tau)|S(\tau)=l}(x)
     \cdot P(c_{m-1} \le D(\tau+\Delta\tau) < c_m \mid D(\tau)=x) \, dx
\end{aligned}
\end{equation}
where $S(\tau)=l$ denotes the event that the system is in CDS $l$ at time $\tau$, and $f_{D(\tau)|S(\tau)=l}(x)$ is the conditional PDF of the damage level at time $\tau$ given it is in state $l$. This conditional PDF is simply the truncated process PDF, $f_{D(\tau)}(x)$, normalized by the total probability of being in that state, $P(S(\tau)=l)$. This leads to the computable formulation:
\begin{equation}
\begin{aligned}
    P_{l \to m}(\tau, \Delta\tau) &= \frac{1}{P(S(\tau)=l)} \int_{c_{l-1}}^{c_l} f_{D(\tau)}(x) \cdot P(c_{m-1} \le D(\tau+\Delta\tau) < c_m | D(\tau)=x) \, dx
\label{eq:integral_prob}
\end{aligned}
\end{equation}
The conditional probability inside the integral can be expressed using the cumulative distribution function (CDF) of the damage increment, $\Delta D = D(\tau+\Delta\tau) - D(\tau)$. The increment $\Delta D$ follows a Gamma distribution with shape $\alpha(\tau+\Delta\tau) - \alpha(\tau)$ and rate $\beta$. Let its CDF be $F_{\Delta D}(\cdot)$. The transition probability becomes:
\begin{equation}
\begin{aligned}
    P_{l \to m}(\tau, \Delta\tau) &= \frac{1}{P(S(\tau)=l)} \int_{c_{l-1}}^{c_l} f_{D(\tau)}(x) \cdot\left[ F_{\Delta D}(c_m - x) - F_{\Delta D}(c_{m-1} - x) \right]\, dx
\end{aligned}
\end{equation}
where $P(S(\tau)=l) = F_{D(\tau)}(c_l) - F_{D(\tau)}(c_{l-1})$. While this integral does not have a general closed-form solution, it can be efficiently and accurately computed using numerical quadrature. This formulation provides a rigorous and physically consistent method for deriving the nonstationary transition matrices, avoiding both the inaccuracies of representative-point approximations and the computational burden of Monte Carlo simulations.

Once Equation~\ref{eq:integral_prob} is evaluated, the pair $(CDS, \tau)$ constitutes the state of the deterioration model by construction and the process is now Markovian by definition. As is generally the case with discretized probabilistic graphical models, the fidelity of the discrete representation is a modeling choice and a closer correspondence to the continuous process can be obtained, when sought, through finer discretization of the relevant states and/or exposure time.

\subsection{Structural Analysis Workflow}
\label{sec:workflow}
\noindent
To assess the structure's performance over its entire design life, a comprehensive simulation workflow is executed, integrating the hazard, deterioration, and structural response models in an iterative Monte Carlo simulation scheme. The structural response is evaluated using nonlinear time history analysis within the OpenSees framework \citep{mazzoni2006opensees}. This workflow is designed to simulate the structural performance across all possible combinations of deterioration levels and hazard intensities. The simulation proceeds in discrete time steps (e.g., annually) over the structure's design life. At each time step $\tau$:
\begin{enumerate}
    \item \textbf{Deterioration Update:} A realization of the corrosion-induced section loss, $D(\tau)$, is sampled from the probabilistic deterioration model (Section~\ref{sec:det}).
    \item \textbf{Model Update:} The finite element analysis (FEA) model in OpenSees is updated to reflect the current state of deterioration. For corrosion, this involves reducing the cross-sectional areas of steel members. The sampled section loss $D(\tau)$ is interpreted as the fractional reduction of the cross-sectional area of the corroding members and is imposed directly on the section geometry, while the material constitutive parameters are left unaltered in this study. This modification directly impacts
the element stiffness and strength matrices, ensuring that
the subsequent structural analysis is state-dependent and
reflects the current physical condition of the structure. More refined effects, such as pitting-induced stress concentrations or ductility loss, can be introduced through corresponding modifications to the section or material models, without affecting the overall framework presented here.  
    \item \textbf{Seismic Event Simulation:} A random check is performed to determine if a seismic event occurs within the time step. This is typically modeled as a Poisson process, where the probability of an event is related to the total mean annual frequency of seismicity at the site.
    \item \textbf{Ground Motion Selection:} If a seismic event is determined to occur, an intensity measure value, $IM$, is sampled from the site's hazard curve, $\lambda(IM)$. A ground motion record is then selected from a pre-defined suite of ground motions, compatible with the sampled $IM$.
    \item \textbf{Nonlinear Dynamic Analysis:} A time history analysis is performed using the updated (deteriorated) structural model and the selected ground motion record. The analysis captures the full nonlinear response of the structure.
    \item \textbf{Record EDPs:} Key EDPs are recorded from the analysis results. These include metrics such as maximum inter-story drifts, roof displacement, and the cumulative dissipated hysteretic energy. These EDPs serve as the primary inputs for the damage model described in the next Section~\ref{sec:damage}.
\end{enumerate}

This structural simulation workflow, as depicted in Figure~\ref{fig:workflow}, providing a clear visual representation of the iterative procedure explained above. This workflow is repeated for a large number of life-cycle realizations to build a robust statistical basis for the generalized fragility analysis. The intensity sampling in this workflow only builds the dataset from which the fragility functions are fitted, and is not by itself the hazard representation of the decision problem, as shown later in the paper. The process is fundamentally path-dependent, as the structural model's state at any time $\tau$ is a function of the entire history of deterioration up to that point, and its response to a seismic event is conditioned on this history. This explicitly couples the slow, continuous process of deterioration with the acute, random process of seismic events, which is a critical feature for realistic life-cycle analysis and a significant departure from traditional, static PEER PBEE assessments \citep{krawinkler20049}.

\begin{table}[tb]
\centering

\caption{Definition of variables for the Park-Ang damage index ($D_{PA}$).}
\label{tab:park_ang_vars}
\begin{tabular}{@{}ll@{}}
\toprule
\textbf{Variable} & \textbf{Description} \\ \midrule
$\delta_m$ & Maximum deformation recorded from the analysis. \\
$\delta_u$ & Ultimate deformation capacity. \\
$\int \mathrm{d}E_h$ & Cumulative dissipated hysteretic energy. \\
$V_y$ & Yield strength of the structure (e.g., yield base shear). \\
$\beta_{PA}$ & Non-negative empirical parameter. \\ \bottomrule
\end{tabular}
\end{table}

\subsection{Damage Analysis}{\label{sec:damage}}
\noindent
To quantify structural damage, particularly the cumulative effects from sequential seismic events, conventional EDPs like maximum inter-story drift are often insufficient. Such metrics do not retain a memory of past loading cycles and are unable to capture damage accumulation from cyclic loading \citep{ibarra2005hysteretic}. Therefore, without loss of generality, we adopt the Park-Ang damage index, $D_{PA}$, a widely recognized and validated metric that combines damage from both maximum deformation and dissipated hysteretic energy \citep{park1985seismic}. This index provides an appropriate measure of the damage state of a structure, especially in a life-cycle context where multiple seismic events of varying intensity may occur \citep{ma2012behavior}. The index is formulated as a linear combination:
\begin{equation}
D_{PA} = \frac{\delta_m}{\delta_u} + \beta_{PA} \frac{\int \mathrm{d}E_h}{V_y \delta_u}
\label{eq:park_ang}
\end{equation}
where the variables are defined in Table~\ref{tab:park_ang_vars}. The first term captures damage from excessive monotonic deformation, while the second term accounts for damage from repeated cyclic loading.

For use in the probabilistic framework, the continuous $D_{PA}$ value calculated after each simulated seismic event is discretized into a finite set of $n_{SDS}$ Seismic Damage States (SDS). These states are defined by $n_{SDS}-1$ thresholds on the $D_{PA}$ scale, typically representing conditions ranging from \textit{Minor Damage} through \textit{Major Damage} to \textit{Collapse}.

\subsection{State-Dependent Generalized Fragility Analysis}
\label{sec:statefrag}
\noindent
This study develops a state-dependent generalized fragility model by extending the generalized fragility framework proposed by Andriotis and Papakonstantinou \citep{andriotis2018extended}. Traditional fragility analysis typically relies on lognormal CDFs to model the probability of exceeding a damage state. This approach can lead to logical inconsistencies, such as crossing fragility curves when dealing with multiple damage states. The extended fragility framework \citep{andriotis2018extended} provides a more rigorous foundation and uses a softmax regression formulation that inherently ensures probabilistic consistency. Furthermore, the generalized fragility framework \citep{andriotis2018extended} defines the IM-conditioned probability of transitioning to a future damage state given the structure's current state, often modeled via dependent Markov formulations.

\begin{figure*}[tb]
    \centering
    \begin{tikzpicture}
        \node[anchor=south west,inner sep=0] (image) at (0,0) {\includegraphics[width=0.7\linewidth]{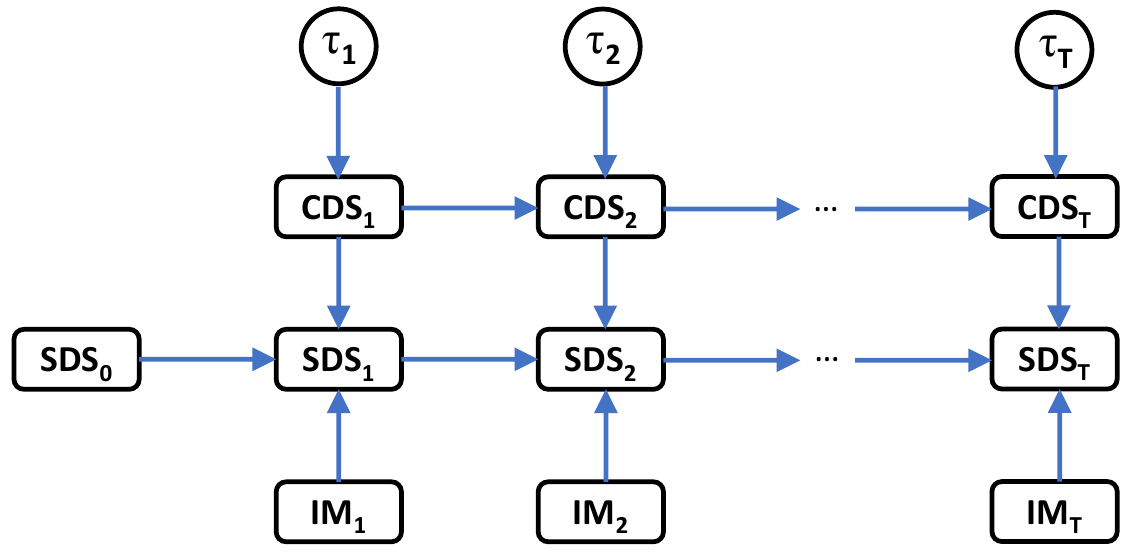}};

        \begin{scope}[x={(image.south east)},y={(image.north west)}]
            \node[anchor=center] at (0.75, 0.7) {\footnotesize $P(CDS_t \mid CDS_{t-1}, \tau_t)$};
            \node[anchor=center] at (0.74, 0.45) {\footnotesize $P(SDS_t \mid SDS_{t-1}, CDS_t, IM_t)$};
        \end{scope}
    \end{tikzpicture}
    \caption{Dynamic Bayesian Network (DBN) for the state-dependent generalized fragility model.}
    \label{fig:dbn}
\end{figure*}

Our work further builds upon this by incorporating an additional layer of conditioning on the underlying structural state into the generalized fragility model. The evolution of seismic damage is formulated as a dependent, discrete-time Markov process, where transition probabilities are conditioned not only on the seismic intensity but also on the structure's evolving state of deterioration and its accumulated damage from prior events. The conditional dependencies of this life-cycle process are concisely represented by the DBN in Figure~\ref{fig:dbn}.

\subsubsection{Dynamic Bayesian Network}
 DBNs are probabilistic graphical models that extend standard Bayesian networks to represent temporal processes \citep{murphy2002dynamic}. They provide an intuitive, graphical representation of the model's assumptions by unrolling the dependencies over discrete time steps. In the DBN, nodes represent random variables and directed edges represent conditional dependencies. The structure formally encodes the Markov property: the future state ($SDS_t$) is independent of all its ancestors given its immediate parents in the graph ($SDS_{t-1}$, $CDS_t$, $IM_t$). This graphical factorization is not merely illustrative, but it provides the formal basis for deriving the joint probability distribution for a sequence of damage events in a tractable manner, as presented in the next section. In particular, the one-step joint transition factorizes here as: 
\begin{equation}
\begin{aligned}
P(CDS_t, SDS_t \mid CDS_{t-1}, SDS_{t-1}, \tau_t, IM_t) 
&= P(CDS_t \mid CDS_{t-1}, \tau_t) \\
&\quad \cdot P(SDS_t \mid SDS_{t-1}, CDS_t, IM_t)
\end{aligned}
\end{equation}
  where the deterioration transition is independent of the seismic damage state and of the intensity, while the seismic transition depends on the deterioration history only through the current corrosion state $CDS_t$. The two transitions are therefore not independent within a time step, and it is precisely this within-slice dependence, encoded by the $CDS_t \rightarrow SDS_t$ edge, that couples the deterioration and seismic damage processes and makes the generalized fragility corrosion-dependent as well. Conversely, the present model assumes the corrosion deterioration evolves independently of the seismic damage state. Interdependencies, such as earthquake-induced damage accelerating subsequent deterioration, can be also represented, if required, by adding an edge from the seismic damage state to the deterioration transition of the DBN, without altering the solution framework.

\subsubsection{Mathematical Formulation}
\noindent
At the core of the formulation is the state transition probability, $p_{j|i,l}(IM_t)$, defined as the probability of the structure transitioning to SDS $j$ during time step $t$, given it was in SDS $i$ and CDS $l$ prior to that, when subjected to intensity $IM_t$:
\begin{equation}
    p_{j|i,l}(im_t) = P(SDS_t = j | SDS_{t-1} = i, CDS_t = l, im_t)
\end{equation}
Based on the conditional independence assumptions of the DBN (Figure~\ref{fig:dbn}), the joint probability mass function for a single damage trajectory $s$ (from a dataset of $N$ simulations) over $T$ time steps can be factored into the product of the one-step transition probabilities. This function also represents the likelihood of observing that single sequence, $\mathcal{L}_s$:
\begin{equation}
\begin{split}
    \mathcal{L}_s &= p(SDS_{1:T}^s | SDS_0^s, CDS_{1:T}^s, IM_{1:T}^s) \\
    &= \prod_{t=1}^{T} p(SDS_t^s | SDS_{t-1}^s, CDS_t^s, im_t^s)
\end{split}
\label{eq:single_likelihood}
\end{equation}

Alternatively, by using an indicator function, $\mathbb{I}(\cdot)$, which equals 1 if its argument is true and 0 otherwise, the likelihood for a single sequence can be written more explicitly by iterating over all possible transitions:
\begin{equation}
\mathcal{L}_s = \prod_{t=1}^{T} \prod_{l=1}^{n_{CDS}} \prod_{i=1}^{n_{SDS}} \prod_{j=1}^{n_{SDS}} p_{j|i,l}(im_t^s)^{\mathbb{I}(SDS_t^s=j, SDS_{t-1}^s=i, CDS_t^s=l)}
\label{eq:single_likelihood_indicator}
\end{equation}

The parameters of the transition probability functions are estimated using the Method of Maximum Likelihood. Assuming the $N$ simulated life-cycle trajectories are independent, the total likelihood of observing the entire dataset, $\mathcal{L}$, is the product of the individual likelihoods:
\begin{equation}
\mathcal{L} = \prod_{s=1}^{N} \mathcal{L}_s = \prod_{s=1}^{N} \prod_{t=1}^{T} p(SDS_t^s | SDS_{t-1}^s, CDS_t^s, im_t^s)
\label{eq:total_likelihood}
\end{equation}

For numerical stability and computational convenience, it is standard practice to maximize the logarithm of the likelihood function. The total log-likelihood, $\ln(\mathcal{L})$, transforms the products into sums. By combining the expressions in Equations~\ref{eq:single_likelihood_indicator} and \ref{eq:total_likelihood} and taking the logarithm, we can express the total log-likelihood by summing over all possible transitions $(i,j,l)$ and selecting the one that occurred at each time step for each simulation:
\begin{equation}
\begin{aligned}
    \ln(\mathcal{L}) &= \sum_{s=1}^{N} \sum_{t=1}^{T} \sum_{l=1}^{n_{CDS}} \sum_{i=1}^{n_{SDS}} \sum_{j=i}^{n_{SDS}} \ln p_{j|i,l}(im_t^s) \cdot \mathbb{I}(\text{SDS}_t^s=j, \text{SDS}_{t-1}^s=i, \text{CDS}_t^s=l)   
\end{aligned}
    \label{eq:log_likelihood_latex}  
\end{equation}

The summation over the final state $j$ starts from the initial state $i$ to enforce the physical constraint that seismic damage is a non-healing process during an event. This log-likelihood function given in Equation~\ref{eq:log_likelihood_latex} serves as the objective function to be maximized by finding the optimal parameters for the transition probability functions. As in the original extended fragility framework, these multi-state transition probabilities are modeled using softmax regression \citep{andriotis2018extended}. 

It is worth noting at this point, that the generality and universality of the proposed framework stems from the way the constituent components are designed, implemented, and integrated. While the hazard curves, fragility functions, and related quantities are hazard- and case-dependent, the overall format of the analysis remains unchanged across hazard types, requiring only the substitution of the corresponding hazard characterization. In this respect, the framework follows the same logic as the PEER PBEE formulation, which was originally introduced for earthquake engineering and has since been adopted for broader risk analyses and other hazard types \citep{barbato2013performance, ciampoli2011performance, lange2014application}.

\section{Life-Cycle Optimization as a Markov Decision Process}
\noindent
This section formally casts the adaptive maintenance problem as a finite-horizon MDP, a powerful mathematical framework for sequential decision-making under uncertainty \citep{puterman2014markov, sutton1998reinforcement}. We first define the MDP components, then highlight the computational challenges posed by system-level optimization, and finally introduce a novel tensor-based algorithm that makes finding the optimal policy tractable. The key notation for the MDP formulation is summarized in Table~\ref{tab:mdp_notation}.

\begin{table*}[tb]
\centering
\caption{Notation for the MDP formulation.}
\label{tab:mdp_notation}
\begin{tabular}{@{}ll@{}}
\toprule
\textbf{Notation} & \textbf{Description} \\ \midrule
$\mathcal{T}$ & Set of decision epochs $\{0, 1, \dots, T-1\}$ over a finite horizon $T$. \\
$\mathcal{S}$, $s$ & Finite set of states, and a specific state $s \in \mathcal{S}$. \\
$\mathcal{A}$, $a$ & Finite set of actions, and a specific action $a \in \mathcal{A}$. \\
$P_a(s'|s, t)$ & Probability of transitioning to state $s'$ from state $s$ under action $a$ at epoch $t$. \\
$C(s, a)$ & Cost incurred for taking action $a$ in state $s$. \\
$\gamma$ & Discount factor for future costs, $0 < \gamma \le 1$. \\
$\pi_t(s)$ & Policy mapping states to actions at epoch $t$, $\pi_t: \mathcal{S} \to \mathcal{A}$. \\
$\pi^*$ & Optimal policy that minimizes total expected discounted cost. \\
$v_t(s)$ & Value (minimum expected future cost) of being in state $s$ at epoch $t$. \\
$N_c$ & Number of components in the infrastructure system. \\
$\mathbf{P}_{SDS}$ & Joint seismic transition matrix over the system seismic damage states. \\
$\mathbf{P}_{SDS|\mathbf{l}}$ & Block of $\mathbf{P}_{SDS}$ for system corrosion state $\mathbf{l}$. \\
$\mathbf{P}_{CDS}^{(k)}$ & Deterioration transition matrix of component $k$. \\
$\mathbf{M}_{a^{(k)}}^{(k)}$ & Maintenance transition matrix of component $k$ under action $a^{(k)}$. \\
$n_{\mathbf{SDS}}$, $n_{\mathbf{CDS}}$ & Number of system seismic and corrosion damage states. \\
$\otimes$ & Kronecker product of matrices. \\
$\mathcal{V}$ & Value function represented as a tensor. \\
$\times_k$ & Mode-$k$ product of a tensor and a matrix. \\ \bottomrule
\end{tabular}
\end{table*}

\subsection{Problem Formulation as a Finite-Horizon MDP}
\label{sec:mdpform}
\noindent
A finite-horizon MDP is defined by the tuple $(\mathcal{T}, \mathcal{S}, \mathcal{A}, P, C, \gamma)$. Each component is specified below in the context of infrastructure life-cycle management.

\textbf{Decision Epochs and Horizon ($\mathcal{T}$):} The infrastructure's life-cycle is discretized into a set of decision epochs, $\mathcal{T} = \{0, 1, \dots, T-1\}$, where each epoch can typically represent one year. Decisions are made at the beginning of each epoch over a finite planning horizon $T$.

\textbf{State Space ($\mathcal{S}$):} The state of the system must encapsulate all necessary information to enable future decisions, satisfying the Markov property \citep{puterman2014markov}. For a single infrastructure component $k$, the state $s^{(k)}$ is a composite variable that includes its seismic damage, deterioration level, and its effective age or exposure time. Specifically, $s^{(k)} = (s_{SDS}^{(k)}, s_{CDS}^{(k)}, s_{\tau}^{(k)})$. The inclusion of the exposure time $\tau$ is a critical modeling choice that allows the nonstationary deterioration process, which depends on the structure's age (as described in Section~\ref{sec:det}), to be captured within a MDP framework. The transition probability from one corrosion state to another depends on this exposure time, which is updated at each step or reset by maintenance actions. For a system of $N_c$ components, the global state space $\mathcal{S}$ is the Cartesian product of the individual component state spaces: $\mathcal{S} = \mathcal{S}^{(1)} \times \mathcal{S}^{(2)} \times \dots \times \mathcal{S}^{(N_c)}$. Consequently, the total number of system states is $|\mathcal{S}| = \prod_{k=1}^{N_c} |\mathcal{S}^{(k)}|$, which grows exponentially with the number of components.

\textbf{Action Space ($\mathcal{A}$):} At each decision epoch, a maintenance action can be chosen for each component. The set of available actions for a component includes options such as Do Nothing, Repair (e.g., improves SDS and CDS by one state and reduces the effective exposure time by 10 years), and Replace (e.g., restores the component to a pristine condition). For a system of $N_c$ components, the system action space $\mathcal{A}$ is the Cartesian product of the component action spaces, $|\mathcal{A}| = \prod_{k=1}^{N_c} |\mathcal{A}^{(k)}|$.

\textbf{Transition Probabilities ($P$):} The state transition probability governs the system's evolution. For the \textit{Do Nothing} action, this probability combines two processes within a time step: gradual deterioration and potential seismic, sudden damage. The probability of transitioning from one combined state $(i,k)$ to another $(j,l)$ (where $i,j$ are SDS indices and $k,l$ are CDS indices) is:
\begin{equation}
\begin{aligned}
    P((j,l)|(i,k)) &= P(SDS_{t+1}=j | SDS_t=i, CDS_{t+1}=l)\cdot P(CDS_{t+1}=l | CDS_t=k, \tau_{t+1})
\end{aligned}
\end{equation}

The deterioration transition probability, $P(CDS_{t+1}=l | CDS_t=k, \tau_{t+1})$, is derived from the nonstationary gamma process as formulated in Section~\ref{sec:det_form}. The seismic transition probability, $P(SDS_{t+1}=j | SDS_t=i, CDS_{t+1}=l)$ in the MDP model, is obtained by marginalizing the state-dependent generalized fragility functions over the full spectrum of seismic hazard uncertainty. This is achieved by integrating over all possible intensity measure values, weighted by their annual probability of occurrence, as derived from the site-specific hazard curves \citep{Clayton2023_NSHMapps, baker2015efficient}:
\begin{equation}
\begin{aligned}
    P(SDS_{t+1}=j | SDS_t=i, CDS_{t+1}=l) &= \int_0^\infty p_{j|i,l}(im) \cdot f_{IM}(im) \, \mathrm{d}im
\end{aligned}
\label{eq:marginal_seismic}
\end{equation}
where $p_{j|i,l}(im)$ is the conditional probability of transitioning to SDS $j$ from SDS $i$ given CDS $l$ and intensity $im$, and $f_{IM}(im)$ is the PDF of the annual seismic intensity, obtained from the hazard curve $\lambda(im)$ as $f_{IM}(im) = \left| d\lambda(im) / d(im) \right|$. Integrating the state-dependent transition probabilities against $f_{IM}(im)$ over the range of relevant intensities corresponds to the probabilistic hazard integration of the PEER PBEE formulation, here carried within the transition model of the decision problem. For a spatially distributed system, the intensities experienced by the individual components during a common seismic event are statistically dependent, as characterized in Section~\ref{sec:hazard}. Since the intensity vector $\mathbf{im}$ is the only quantity shared across components, the seismic transitions remain conditionally independent given $\mathbf{im} = (im^{(1)}, \dots, im^{(N_c)})$, and the joint transition of the system seismic damage vector $\mathbf{SDS} = (SDS^{(1)}, \dots, SDS^{(N_c)})$ is obtained by marginalizing the product of the component transition probabilities over the joint intensity distribution:
\begin{equation}
\begin{aligned}
    &P(\mathbf{SDS}_{t+1} = \mathbf{j} \mid \mathbf{SDS}_t = \mathbf{i},
      \mathbf{CDS}_{t+1} = \mathbf{l}) =  \int_{\mathbb{R}^{N_c}} \prod_{k=1}^{N_c}
      p_{j_k | i_k, l_k}^{(k)}\!\left(im^{(k)}\right)
      f_{\mathbf{IM}}(\mathbf{im}) \, \mathrm{d}\mathbf{im}
\end{aligned}
\label{eq:joint_seismic}
\end{equation}
where $f_{\mathbf{IM}}(\mathbf{im})$ is the joint annual density of the intensity vector, obtained from the site-specific hazard curves and the correlated field of Section~\ref{sec:hazard}. When the sites are treated as independent, $f_{\mathbf{IM}}$ factorizes and Equation~\ref{eq:joint_seismic} reduces to the product of the component-wise marginals of Equation~\ref{eq:marginal_seismic}. Transition probabilities for other actions, such as Repair or Replace, are modeled in this work as deterministic shifts to lesser damaged states. These shifts act on the full component state, including the exposure time $\tau$, so that the subsequent deterioration dynamics are updated automatically through the $\tau$-dependence of the nonstationary transition probabilities of Section~\ref{sec:det}, with a repaired component evolving thereafter as a less-aged structure, and a replaced one from the as-built condition. Within each epoch, the maintenance action is applied first, and the resulting condition is then exposed to the deterioration and seismic transitions above, consistent with decisions being taken at the beginning of the epoch.

\textbf{Cost Function and Discount Factor ($C, \gamma$):} The cost function $C(s, a)$ includes the immediate cost of implementing action $a$ and any state-dependent costs associated with being in state $s$ (e.g., costs associated with partial or full failure). More broadly, $C(s,a)$ is not restricted to monetary cost. Any objective of interest, such as downtime, or sustainability-related measures, or a combination thereof, can be encoded in the same term without any change to the formulation or the solution method.  Future costs are discounted by a factor $\gamma$ to reflect the time value of money. The objective is to find an optimal policy $\pi^*$ that minimizes the total expected discounted life-cycle cost.

\subsection{Computational Challenge: The Curse of Dimensionality}
\noindent
The optimal policy $\pi^*$ can be found using dynamic programming \citep{puterman2014markov, sutton1998reinforcement}, which solves the Bellman optimality equation to find the minimum expected cost/utility, or value $v_t^*(s)$, for each state $s$ at each time step $t$:
\begin{equation} \label{eq:bellman}
   v_t^*(s) = \min_{a \in \mathcal{A}} \left\{ C(s, a) + \gamma \sum_{s' \in \mathcal{S}} P(s' | s, a, t) v_{t+1}^*(s') \right\}
\end{equation}
This is solved via backward induction, starting from $v_T^*(s) = 0, \forall s \in \mathcal{S}$. However, its direct application is computationally intractable for system-level problems due to the curse of dimensionality \citep{bertsekas2012dynamic, sutton1998reinforcement}. The state and action spaces, $|\mathcal{S}|$ and $|\mathcal{A}|$, grow exponentially with the number of components $N_c$. Since the complexity of each backward step is roughly $\mathcal{O}(|\mathcal{S}|^2 \cdot |\mathcal{A}|)$, the computational and memory requirements become prohibitive for even moderately sized systems, rendering standard solvers infeasible. This challenge necessitates a more scalable solution approach.

\subsection{Efficient Solution via Tensor-Based Value Iteration}\label{sec:tensormdp}
\noindent
To mitigate the curse of dimensionality, we introduce a novel computational method that exploits the inherent structure of the problem, which is categorized as a factored MDP. Unlike existing factored solution algorithms that rely on approximate value functions and linear basis combinations to maintain tractability, our method utilizes tensor algebra and Kronecker products to solve the MDP exactly \citep{degris2013factored, guestrin2003efficient}. This allows for an exponential reduction in computational complexity without loss of accuracy or accumulation of errors that can be found in approximate factored schemes. While DRL has proven very effective for navigating the curse of dimensionality \citep{andriotis2019managing, saifullah2026multi}, this work focuses on an exact solution approach to prioritize global optimality. The developed framework, however, fully supports DRL as an alternative approximate solver.

\subsubsection{Kronecker Product Representation}\label{sec:kron}
\noindent
The key insight is that for many infrastructure systems, the physical evolution of one component is conditionally independent of the state of other components, given a maintenance action. This assumption is not overly restrictive, as statistical independence can be maintained conditional on common hyperparameters within the DBN \citep{luque2019risk, morato2023inference}. This conditional independence is explicitly encoded in our DBN model (Figure~\ref{fig:dbn}) and allows the system's transition probability matrix $\mathbf{P}_a$ for a joint action $a = (a^{(1)}, \dots, a^{(N_c)})$ to be represented as the Kronecker product of the individual component transition matrices $\mathbf{P}_{a^{(k)}}^{(k)}$ \citep{davio2012kronecker, steeb2011matrix}:
\begin{equation}
    \mathbf{P}_a = \mathbf{P}_{a^{(1)}}^{(1)} \otimes \mathbf{P}_{a^{(2)}}^{(2)} \otimes \dots \otimes \mathbf{P}_{a^{(N_c)}}^{(N_c)}
    \label{eq:kronecker}
\end{equation}
This representation avoids the explicit construction and storage of the enormous system transition matrix $\mathbf{P}$. Instead, only the smaller component-level matrices are needed. This condition is not fully met, however, when the components are exposed to a spatially correlated hazard, as in Equation~\ref{eq:joint_seismic}. Yet, the independence of the components conditional on the shared hazard variables is still preserved, so that the factored representation is only partially, rather than entirely, lost. Equation~\ref{eq:kronecker} is retained throughout as the fully separable case, and the corresponding generalization is developed in Section~\ref{sec:coupled}. 

\subsubsection{Tensor-Based Value Iteration Update}
\label{sec:tensorupdate}
\noindent
With the transition dynamics represented by Kronecker products, the value iteration update can be performed efficiently using tensor operations \citep{kolda2009tensor, rabanser2017introduction}. The core of each update step involves calculating the expected future value, $\sum_{s'} P_a(s'|s) v_{t+1}(s')$, for each state $s$ and action $a$. In vector notation, this corresponds to the matrix-vector product $\mathbf{P}_a \mathbf{v}_{t+1}$. Substituting the Kronecker representation from Equation~\ref{eq:kronecker} allows us to write this as:
\begin{equation}
    \mathbf{P}_a \mathbf{v}_{t+1} = \left( \mathbf{P}_{a^{(1)}}^{(1)} \otimes \mathbf{P}_{a^{(2)}}^{(2)} \otimes \dots \otimes \mathbf{P}_{a^{(N_c)}}^{(N_c)} \right) \mathbf{v}_{t+1}
    \label{eq:kronecker_product_vector}
\end{equation}

While this equation still involves the conceptual full-system matrix, its structure can be exploited to avoid its explicit formation. This is achieved by leveraging the mixed-product property of the Kronecker product, which allows the large matrix-vector product to be recast as a sequence of much smaller tensor-matrix operations. To do this, the value vector $\mathbf{v}_{t+1} \in \mathbb{R}^{|\mathcal{S}|}$ is first reshaped into an $N_c$-order tensor $\mathcal{V}_{t+1} \in \mathbb{R}^{|\mathcal{S}^{(1)}| \times \dots \times |\mathcal{S}^{(N_c)}|}$. The product in Equation~\ref{eq:kronecker_product_vector} is then computed via a series of mode-wise products, where $\times_k$ denotes the mode-$k$ product:
\begin{align}
    \mathcal{V}^{(1)} &= \mathcal{V}_{t+1} \times_1 \mathbf{P}_{a^{(1)}}^{(1)} \label{eq:mode1_prod} \\
    \mathcal{V}^{(2)} &= \mathcal{V}^{(1)} \times_2 \mathbf{P}_{a^{(2)}}^{(2)} \label{eq:mode2_prod} \\
    &\vdots \nonumber \\
    \mathcal{V}^{(N_c)} &= \mathcal{V}^{(N_c-1)} \times_{N_c} \mathbf{P}_{a^{(N_c)}}^{(N_c)} \label{eq:moden_prod}
\end{align}
The final tensor, $\mathcal{V}^{(N_c)}$, contains the expected future values, and vectorizing it would yield the exact result of the original product, i.e., $\text{vec}(\mathcal{V}^{(N_c)}) = \mathbf{P}_a \mathbf{v}_{t+1}$. By decomposing the most computationally intensive step, the proposed tensor-based approach reduces the update complexity from exponential to linear in the number of components, for separable dynamics. This separable update is generalized to spatially correlated hazards in Section~\ref{sec:coupled}, where the complete procedure, together with a comparison of its computational cost against naive and sparse solvers, are presented.

\begin{algorithm}[tb]
\caption{Tensor-Based Value Iteration for MDPs}
\label{alg:tensor_vi}
\begin{algorithmic}[1]
\REQUIRE Maintenance matrices $\{\mathbf{M}_{a^{(k)}}^{(k)}\}$ and deterioration matrices
$\{\mathbf{P}_{CDS}^{(k)}\}$ for each component $k \in \{1, \dots, N_c\}$ and action
$a^{(k)} \in \mathcal{A}^{(k)}$; joint seismic transition matrix $\mathbf{P}_{SDS}$; cost
function $C(s, a)$; discount factor $\gamma$; horizon $T$.
\STATE Set mode matrices $\mathbf{A}_{a^{(k)}}^{(k)} \gets \mathbf{M}_{a^{(k)}}^{(k)}\mathbf{P}_{CDS}^{(k)}$; for the separable case set $\mathbf{A}_{a^{(k)}}^{(k)} \gets \mathbf{P}_{a^{(k)}}^{(k)}$ (Eq.~\ref{eq:kronecker}), which already carries the $\mathbf{P}_{SDS}^{(k)}$, so that the joint update of Line~6 gives $\mathcal{V}^{(0)} = \mathcal{V}_{t+1}$.
\ENSURE Optimal value function $\mathbf{v}^*$; optimal policy $\pi^*$
\STATE Initialize terminal value: $v_T(s)$ for all $s \in \mathcal{S}$
\FOR{$t = T-1$ \textbf{down to} $0$}
    \FOR{each system action $a = (a^{(1)}, \dots, a^{(N_c)}) \in \mathcal{A}$}
        \STATE Reshape $\mathbf{v}_{t+1} \in \mathbb{R}^{|\mathcal{S}|}$ into tensor $\mathcal{V}_{t+1} \in \mathbb{R}^{|\mathcal{S}^{(1)}| \times \cdots \times |\mathcal{S}^{(N_c)}|}$
        \STATE $\mathcal{V}^{(0)} \gets \mathbf{P}_{SDS} \, \mathcal{V}_{t+1}$ \COMMENT{Joint seismic hazard update}
        \FOR{$k = 1$ \textbf{to} $N_c$}
            \STATE $\mathcal{V}^{(k)} \gets \mathcal{V}^{(k-1)} \times_k \mathbf{A}_{a^{(k)}}^{(k)}$ \COMMENT{Mode-$k$ product}
        \ENDFOR
        \STATE $\mathbf{v}_a \gets \text{vec}(\mathcal{V}^{(N_c)})$
        \STATE $q_t(\cdot, a) \gets C(\cdot, a) + \gamma \, \mathbf{v}_a$
    \ENDFOR
    \FOR{each state $s \in \mathcal{S}$}
        \STATE $v_t^*(s) \gets \min_{a \in \mathcal{A}} q_t(s, a)$
        \STATE $\pi_t^*(s) \gets \arg\min_{a \in \mathcal{A}} q_t(s, a)$
    \ENDFOR
\ENDFOR
\RETURN $\mathbf{v}^*$, $\pi^*$
\end{algorithmic}
\end{algorithm}

\subsubsection{Generalization to Partially Coupled Transition Structures}
\label{sec:coupled}
\noindent
The Kronecker representation of Equation~\ref{eq:kronecker} presumes that the one-step evolution of each component is independent of the others given the applied action. Spatially correlated hazards violate this premise, as described by Equation~\ref{eq:joint_seismic}. This dependence, however, does not destroy the factored structure of the problem, since it is confined to the seismic damage states alone and leaves the remaining dynamics separable. The tensor-based update is therefore retained, with the coupling treated exactly on a low-dimensional subspace.

To formalize this, the system transition matrix is now decomposed by physical mechanisms rather than by components alone. Denoting by $\mathbf{M}_{a^{(k)}}^{(k)}$ the deterministic transition associated with the maintenance action $a^{(k)}$, by $\mathbf{P}_{CDS}^{(k)}$ the deterioration transition matrix of Section~\ref{sec:det}, and by $\mathbf{P}_{SDS}$ the joint seismic transition matrix of Equation~\ref{eq:joint_seismic}, the system transition matrix now reads:
\begin{equation}
    \mathbf{P}_a = \left( \bigotimes_{k=1}^{N_c} \mathbf{M}_{a^{(k)}}^{(k)} \mathbf{P}_{CDS}^{(k)} \right) \mathbf{P}_{SDS}
    \label{eq:split}
\end{equation}
where the factors are ordered as they occur within the epoch, that is, the maintenance intervention, then the deterioration over the year, then the exposure to the seismic hazard. The entire departure from Equation~\ref{eq:kronecker} is contained in $\mathbf{P}_{SDS}$. When the sites are independent, $\mathbf{P}_{SDS}$ factorizes as $\bigotimes_{k} \mathbf{P}_{SDS}^{(k)}$, where $\mathbf{P}_{SDS}^{(k)}$ is the marginalized seismic transition of component $k$ given by Equation~\ref{eq:marginal_seismic}. The mixed-product property then gives $\mathbf{P}_{a^{(k)}}^{(k)} = \mathbf{M}_{a^{(k)}}^{(k)} \mathbf{P}_{CDS}^{(k)} \mathbf{P}_{SDS}^{(k)}$, recovering the logic of Section~\ref{sec:kron}.

\begin{figure*}[!tb]
    \centering
    \includegraphics[width=0.9\textwidth]{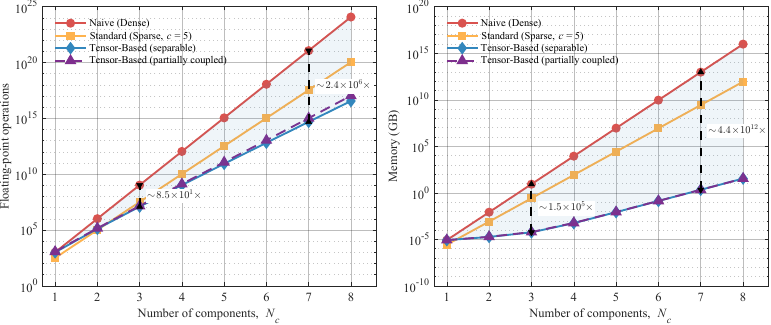}
    
    \makebox[0.14\textwidth][c]{(a)} \hspace{0.20\textwidth} \makebox[0.14\textwidth][c]{(b)}
    
\caption{\textcolor{black}{Computational complexity comparison for a homogeneous system with
$|\mathcal{S}^{(k)}| = 16$ states and $|\mathcal{A}^{(k)}| = 4$ actions per component,
each component comprising $n_{SDS}^{(k)} = 2$ and
$n_{CDS}^{(k)} = 2$ over the exposure-time levels, and
$c = 5$ average non-zeros per row per component matrix. (a)~Total floating-point
operations per value iteration update. (b)~Total memory requirement in GB.
Vertical dashed lines indicate the ratio between the naive and
separable tensor-based methods at selected values of $N_c$.}}
    \label{fig:complexity_comparison}
\end{figure*}

\begin{table*}[tb]
\centering
\small
\caption{Comparison of computational complexities for a single value iteration update. $nnz(\mathbf{P}_a)$ denotes the number of nonzero entries in the system transition matrix.}
\label{tab:complexity_comparison}
\begin{tabular}{@{}lll@{}}
\toprule
\textbf{Method} & \textbf{Memory Complexity} & \textbf{Time Complexity} \\ \midrule
Naive (Dense) & $\mathcal{O}(|\mathcal{A}||\mathcal{S}|^2+ |\mathcal{S}|)$ & $\mathcal{O}(|\mathcal{A}||\mathcal{S}|^2)$ \\
Standard (Sparse) & $\mathcal{O}(|\mathcal{A}| \cdot nnz(\mathbf{P}_a) + |\mathcal{S}|)$ & $\mathcal{O}(|\mathcal{A}| \cdot nnz(\mathbf{P}_a))$ \\
Tensor-Based (separable) & $\mathcal{O}(\sum_{k=1}^{N_c} |\mathcal{A}^{(k)}| |\mathcal{S}^{(k)}|^2 + |\mathcal{S}|)$ & $\mathcal{O}(|\mathcal{A}||\mathcal{S}| \sum_{k=1}^{N_c} |\mathcal{S}^{(k)}|)$ \\ 
Tensor-Based (coupled) &
$\mathcal{O}(\sum_{k=1}^{N_c} |\mathcal{A}^{(k)}| |\mathcal{S}^{(k)}|^2
 + n_{\mathbf{CDS}} \, n_{\mathbf{SDS}}^2 + |\mathcal{S}|)$ &
$\mathcal{O}(|\mathcal{A}||\mathcal{S}|
 (\sum_{k=1}^{N_c} |\mathcal{S}^{(k)}| + n_{\mathbf{SDS}}))$ \\ \bottomrule
\end{tabular}
\end{table*}

A seismic event changes only the seismic damage of the components; it leaves the corrosion states and exposure times unchanged in this case. Writing a system state as $(\mathbf{SDS}, \mathbf{CDS}, \boldsymbol{\tau})$, the joint seismic transition therefore acts only on $\mathbf{SDS}$, and is conditioned on the corrosion configuration $\mathbf{l} = \mathbf{CDS}$ through the state-dependent fragility:
\begin{equation}
   P_{SDS}\big( (\mathbf{j}, \mathbf{l}, \boldsymbol{\tau}) \mid (\mathbf{i}, \mathbf{l}, \boldsymbol{\tau}) \big)
   = P_{SDS|\mathbf{l}}(\mathbf{j} \mid \mathbf{i}),
   \quad \mathbf{P}_{SDS|\mathbf{l}} \in \mathbb{R}^{n_{\mathbf{SDS}} \times n_{\mathbf{SDS}}}
   \label{eq:psds}
\end{equation}
with all transitions that would alter $\mathbf{CDS}$ or $\boldsymbol{\tau}$ being zero. Here $\mathbf{i}$ and $\mathbf{j}$ are the system seismic damage states before and after the event, $n_{\mathbf{SDS}} = \prod_{k} n_{SDS}^{(k)}$ is the number of such states, and there is one block $\mathbf{P}_{SDS|\mathbf{l}}$ for each of the $n_{\mathbf{CDS}} = \prod_{k} n_{CDS}^{(k)}$ corrosion configurations. The joint seismic transition is therefore fully described by $n_{\mathbf{CDS}} \, n_{\mathbf{SDS}}^{2}$ entries, independent of the exposure-time resolution.

Because the intensity vector is the only variable shared among the components, $\mathbf{P}_{SDS}$ also admits the equivalent representation:
\begin{equation}
    \mathbf{P}_{SDS} = \mathbb{E}_{\mathbf{IM}} \!\left[
    \bigotimes_{k=1}^{N_c} \mathbf{P}_{SDS}^{(k)}\!\left(im^{(k)}\right) \right]
    \label{eq:psds_mixture}
\end{equation}
where $\mathbf{P}_{SDS}^{(k)}(im^{(k)})$ is the intensity-conditioned seismic transition matrix of component $k$, and the expectation is taken over the joint annual intensity distribution of Equation~\ref{eq:joint_seismic}, including the outcome of no event, for which each factor reduces to the identity. Equation~\ref{eq:psds_mixture} indicates conditional independence given the shared hazard variables in Section~\ref{sec:kron}, showing the joint seismic transition matrix to be a mixture of Kronecker products. The expectation in Equation~\ref{eq:psds_mixture}, and the relevant intensity marginalizations in Equations~\ref{eq:marginal_seismic}--\ref{eq:joint_seismic}, are evaluated in this work by $10^{6}$ Monte Carlo samples, based also on the correlated field presented in Section~\ref{sec:hazard}.

The order of Equation~\ref{eq:split} must be respected, since $\mathbf{P}_{SDS}$ does not commute with the separable factors and a different order would correspond to a different model of the epoch. Both the fully separable representation of Equation~\ref{eq:kronecker} and the partially coupled representation of Equation~\ref{eq:split} are solved by a single tensor-based value iteration, summarized in Algorithm~\ref{alg:tensor_vi}. Substituting Equation~\ref{eq:split} into the Bellman update, the expected future value is obtained by first contracting the value tensor with $\mathbf{P}_{SDS}$ and then performing the mode-wise products of Equations~\ref{eq:mode1_prod}--\ref{eq:moden_prod} with the separable factors. For fully separable dynamics, this joint seismic update (Line~6 of Algorithm~\ref{alg:tensor_vi}) is omitted and the mode-$k$ matrices reduce to the component transition matrices $\mathbf{P}_{a^{(k)}}^{(k)}$ of Equation~\ref{eq:kronecker}. In either case the solution remains exact, and the system transition matrix is at no point explicitly formed.

In practice, the loop over system actions (Line 4) can be parallelized across actions, and the mode-$k$ products (Lines 7--9) can leverage optimized tensor libraries. When the maintenance transitions are deterministic shifts, $\mathbf{M}_{a^{(k)}}^{(k)}$ reduces to an index gather, that is, a direct reindexing of the value tensor entries rather than a matrix multiplication, and the action-independent part of Equation~\ref{eq:split} can be evaluated once per epoch and reused across all system actions. The resulting update cost outperforms naive dynamic programming methods and sparse matrix formulations, such as the one in the efficient MATLAB \texttt{MDPToolbox} \citep{chades2014mdptoolbox}, as shown in Table~\ref{tab:complexity_comparison} and Figure~\ref{fig:complexity_comparison}.

To illustrate the practical impact of these complexity orders in Table~\ref{tab:complexity_comparison}, Figure~\ref{fig:complexity_comparison} highlights the time and memory complexities of the evaluated methods as a function of the component count $N_c$ for a homogeneous system defined by $|\mathcal{S}^{(k)}| = 16$, $|\mathcal{A}^{(k)}| = 4$, and $c = 5$ non-zero transition probabilities per row. By operating strictly on local component matrices and avoiding the explicit construction of exponentially sized system-level transition matrices, the tensor method achieves substantial computational speedups and memory reductions relative to both the naive and sparse baselines. This linear scaling applies to fully separable dynamics. The partially coupled case retains one additional term, associated with the joint transition on the spatially dependent subspace, here manifested through seismic damage dependence. This leaves the memory requirement essentially unchanged and adds only a slight overhead in operation count, remaining far below the naive and sparse solver costs as the system grows.

\section{Numerical Example}
\label{sec:numerical_example}
\noindent
This section demonstrates the framework's application to a critical network in a high-seismicity urban environment. The case study is designed to be complex enough to highlight challenges like component inter-dependencies and the curse of dimensionality while remaining tractable to clearly illustrate the methodology.

\subsection{System Definition and Scenario}
\begin{figure}[tb]
    \centering
    \begin{subfigure}[c]{0.45\textwidth}
        \centering
        \includegraphics[width=\textwidth]{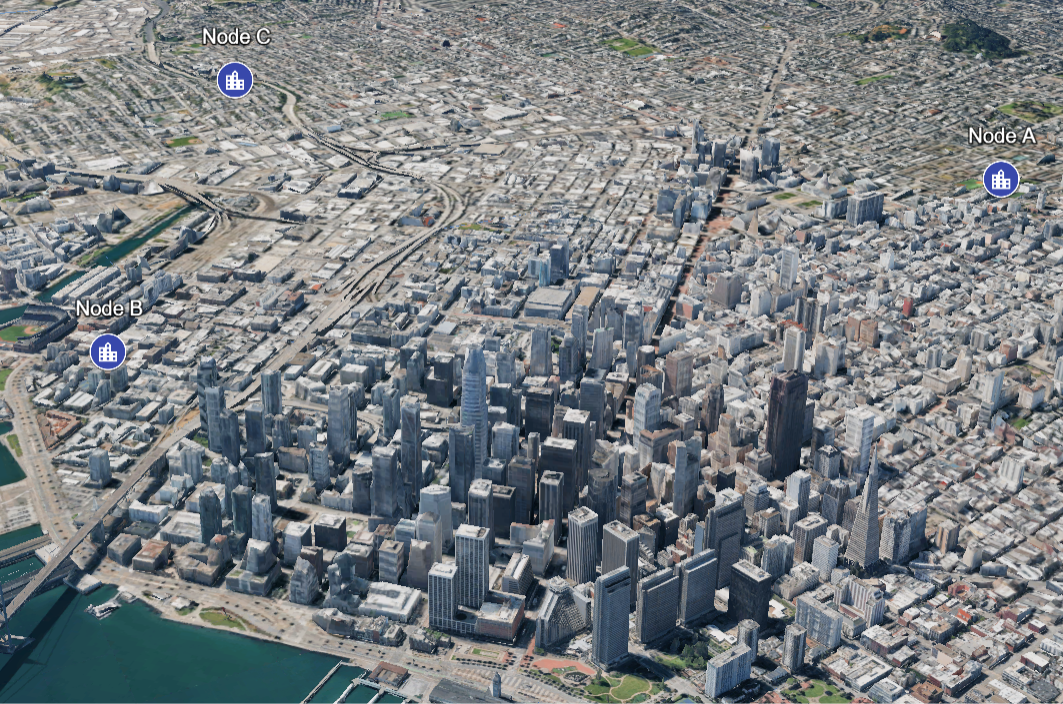}
        \caption{Spatial layout of the three system nodes in San Francisco, CA (37.7818°N, 122.4020°W) from Google Earth Web. Map data: Google; Imagery ©2025 Maxar Technologies (as displayed on image). Accessed 14 April 2026. \citep{GoogleEarthPro2025}}
        \label{fig:map_sub}
    \end{subfigure}
    \hfill 
    \begin{subfigure}[c]{0.4\textwidth}
        \centering
        \vspace{0.5cm}
        \begin{tikzpicture}[node distance=2.5cm, thick, scale=0.8, every node/.style={transform shape}, >={Stealth[scale=1.2]}]
    \node (power) [rectangle, draw, align=center] {Node\\A};
    \node (dec) [rectangle, draw, below of=power, xshift=-3cm, align=center] {Node\\B};
    \node (firehq) [rectangle, draw, below of=power, xshift=3cm, align=center] {Node\\ C};

    \draw[->] (dec) -- node[left] {} (power);
    \draw[->] (firehq) -- node[right] {} (power);
    \draw[->] (firehq) -- node[below] {} (dec);
\end{tikzpicture}
        \caption{Dependency diagram.}
        \label{fig:dependency_sub}
    \end{subfigure}
    
    \caption{Location and dependency of three infrastructure nodes. (a) The geographical location of nodes A, B, and C in San Francisco, CA. (b) The corresponding system dependency diagram.}
    \label{fig:sf_system}
\end{figure}
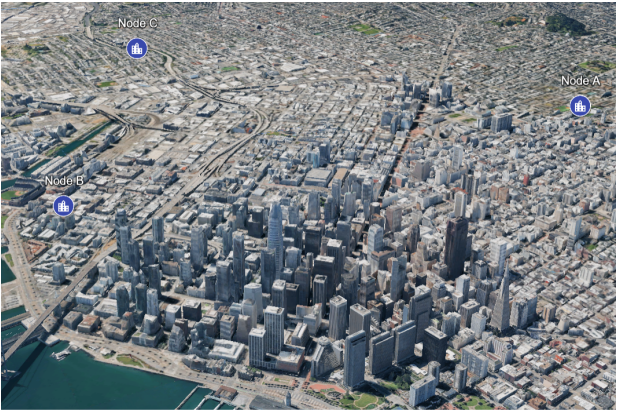
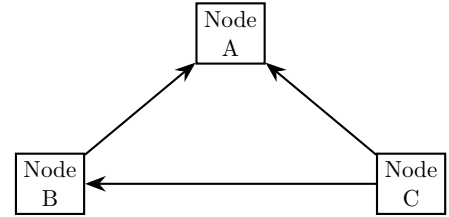

\noindent
We consider a hypothetical but plausible emergency response infrastructure network evaluated over a 50-year design life in San Francisco, California. As depicted in Figure~\ref{fig:sf_system}, the network comprises three functionally interdependent nodes critical for disaster response. For illustrative purposes, Node A can be reasonably interpreted as a hospital, while Nodes B and C can represent a water pumping station and an electrical substation, respectively. These nodes form a cascading dependency structure wherein the substation (C) provides power to the water station (B), while the hospital (A) requires both utilities to remain operational. Without loss of generality, Figure~\ref{fig:sf_system} can serve as a representative reliability block diagram, capturing the fundamental logic of functional dependencies inherent in larger, more complex infrastructure systems. Each node is modeled as a representative steel moment-resisting frame building: a two-story, four-bay structure (A), a three-story, three-bay structure (B), and a four-story, two-bay structure (C). \textcolor{black}{A site-specific seismic hazard curve is used for each of the
three nodes, obtained at coordinates (37.7860$^\circ$N, 122.4217$^\circ$W), (37.7826$^\circ$N, 122.3909$^\circ$W), and (37.7604$^\circ$N, 122.4031$^\circ$W) for Nodes A, B, and C, respectively, under a common reference site condition of $V_{s30} = 760$~m/s.}

\begin{table}[tb]
\centering
\caption{Structural analysis parameters.}
\label{tab:structural_params}
    \begin{tabular}{l l}
        \hline
        \textbf{Parameter} & \textbf{Value} \\
        \hline
        Material yield strength & 235 MPa \\
        Elastic modulus & 200 GPa \\
        Strain hardening ratio & 0.5\% \\
        Beam section & W 21x83 \\
        Column section & W 24x84 \\
        Concrete slab thickness & 0.3 m \\
        Beam length & 6.5 m \\ 
        Column length & 4.2 m \\
        \hline
    \end{tabular}
\end{table}

\begin{table*}[tb]
\centering
\caption{System-level failure scenarios and associated costs.}
\label{tab:failure_costs}
\begin{tabular}{@{} l c c c @{}}
\toprule
\textbf{Failure Scenario} & \textbf{Direct Cost} & \textbf{I-O Impact Multiplier} & \textbf{Total System Cost} \\ 
\midrule
A Fails          & \$423k   & 1.3 & \$0.55M \\
B Fails          & \$684k   & 1.4 & \$0.96M \\
C Fails          & \$793k   & 1.5 & \$1.19M \\
A \& C Fail     & \$1,216k & 2.0 & \$2.43M \\
B \& C Fail     & \$1,477k & 2.2 & \$3.25M \\
A \& B Fail     & \$1,107k & 1.8 & \$1.99M \\
A, B, \& C Fail & \$1,900k & 3.0 & \$5.70M \\ 
\bottomrule
\end{tabular}
\end{table*}
To evaluate the seismic performance of these archetypes, a 2D nonlinear structural model was developed for each in OpenSees Version 3.7.0 \citep{mckenna2011opensees}. The structural members were modeled using fiber-discretized elements, which capture the spread of inelasticity both along the member length and across the section depth, allowing an accurate simulation of material nonlinearity under cyclic loading. The steel material was modeled using an isotropic strain hardening law. Each member is modeled with five integration points and fully fixed bases, and the seismic mass is lumped at the nodes from the tributary gravity loads. Rayleigh damping with a 5\% ratio is assigned, and the nonlinear response-history analysis employs the Newmark average-acceleration scheme with a 0.01~s time step and a convergence criterion with tolerance of $10^{-3}$. Other key parameters used in the structural analysis are summarized in Table~\ref{tab:structural_params}. The synthetic ground motions are drawn from a suite of 12{,}000 non-stationary records generated using the predictive stochastic model of~\cite{vlachos2016multi, vlachos2018predictive}, simulated by sampling the magnitude and epicentral distance uniformly over $M \in [5.5, 8.0]$ and $R \in [5.0, 15.0]$~km. The Park-Ang model parameters $\delta_u$ and $V_y$ in Eq.~\eqref{eq:park_ang} are obtained from a static pushover analysis of each frame archetype, with $\beta_{PA}=0.025$.

\begin{table}[tb]
\centering
\caption{CDS definitions.}
\label{tab:cds_def}
\begin{tabular}{@{}ccl@{}}
\toprule
\textbf{CDS} & \textbf{Section Loss} & \textbf{Description} \\ \midrule
1            & $<$ 0.1                    & Negligible deterioration. \\
2            & 0.1 -- 0.4                 & Minor to moderate deterioration.  \\
3            & $>$ 0.4                    & Significant deterioration.  \\ \bottomrule
\end{tabular}
\end{table}
\begin{table}[tb]
\centering
\caption{SDS definitions.}
\label{tab:sds_def}
\begin{tabular}{@{}ccl@{}}
\toprule
\textbf{SDS} & \textbf{$D_{PA}$ Range} & \textbf{Description} \\ \midrule
1            & $<$ 0.2                 & No damage or minor, cosmetic damage. \\
2            & 0.2 -- 0.5              & Moderate, repairable structural damage.\\
3            & $>$ 0.5                 & Severe damage or collapse.\\ \bottomrule
\end{tabular}
\end{table}
A crucial aspect of this example is the modeling of system-level interdependencies. The economic consequences of component failure extend beyond the direct repair or replacement cost of an individual node. The functional dependency among the nodes means that the failure of one can have cascading effects, amplifying the impact on the system's overall operational capacity. To estimate the broader economic impacts of service disruptions, we use an input–output approach aligned with FEMA's Hazus methodology \citep{fema2022hazus}. Specifically, we draw on the Regional Input-Output Modeling System (RIMS II), a tool commonly used by policymakers, planners, and investors to evaluate regional economic impacts \citep{bess2011input}. RIMS II provides multipliers that quantify the ripple effects of disruptions across sectors. In lieu of a purchased dataset, we use illustrative values consistent with the San Francisco Bay Area, based on published RIMS II multipliers \citep{chsra2020rims}. The total system-level cost for any given failure scenario is thus the sum of the direct costs of the failed components multiplied by the corresponding multiplier. The resulting failure cost is therefore non-additive in the component states, with $C_F(s) > \sum_{k} C_F^{(k)}(s^{(k)})$ whenever more than one component is in its failure state, and it is through this term that the functional dependencies of Figure~\ref{fig:dependency_sub} enter the optimization. The components are thus coupled in the objective through the cost, and in the dynamics through the correlated seismic hazard transition of Section~\ref{sec:coupled}. The direct costs for component are sourced from the NHERI SimCenter's R2D tool \citep{McKenna2025R2DTool, Deierlein2020Cloud}. This cost structure creates complex trade-offs for the optimization algorithm; for instance, the optimal policy may prioritize protecting a specific node not only due to its high direct failure cost but also because its failure significantly exacerbates the consequences of other potential failures. The failure costs and associated I-O Impact Multiplier for this example are detailed in Table~\ref{tab:failure_costs}.

\paragraph{Deterioration Model:}
As discussed earlier the gradual degradation of the steel frame components is modeled as corrosion-induced section loss. For this specific case we employ the power-law shape function, $\alpha(\tau) = a\tau^b$, with an exponent of $b=1.5$ to capture the non-linear behavior. The parameter $a$ of Equation~\ref{eq:power_law} and the rate parameter $\beta$ of Equation~\ref{eq:gamma_pdf} are calibrated such that the model approximates a mean section loss of 0.4 and results in a standard deviation, $\sigma$, of 7.5\% after 50 years, yielding $a \approx 0.0805$ and $\beta \approx 71.1$, representing sufficient long-term structural uncertainty. This physical process is categorized into three mutually exclusive and collectively exhaustive CDS, as defined in Table~\ref{tab:cds_def}. This discretization is a necessary step to construct a finite state space for the MDP, aligning with the nonstationary gamma process formulation detailed in Section~\ref{sec:det}. The number of states and their thresholds are selected to represent physically distinct, actionable condition levels, i.e., negligible, moderate, and significant deterioration, consistent with common condition-rating practice, while keeping the state space finite and tractable. A finer discretization increases fidelity at the expense of a larger state space, and can be selected following dedicated studies of state-discretization resolution~\citep{morato2022optimal} when a closer representation of the continuous deterioration dynamics is sought.

\paragraph{Seismic Damage Model:}
As mentioned earlier, damage incurred during a seismic event is quantified using the Park-Ang damage index ($D_{PA}$), as formulated in Equation~\ref{eq:park_ang}. The continuous $D_{PA}$ values obtained from structural analysis are discretized into three SDS, as defined in Table~\ref{tab:sds_def}. These thresholds are selected based on calibrated damage scales from established literature \citep{park1985mechanistic, park1987damage}, where SDS 3 ($D_{PA} \ge 0.5$) is explicitly defined as the component failure state. While theoretical collapse corresponds to $D_{PA} \ge 1.0$, the threshold of 0.5 is adopted here as a conservative lower bound representing the onset of severe, irreparable damage \citep{park1987damage}. When a component transitions into this state, the corresponding failure costs from Table~\ref{tab:failure_costs} are triggered within the MDP's reward model.

\subsection{MDP Specification and Computational Challenge}
\noindent
For this analysis, each component is modeled with three discrete damage states for both corrosion and seismic performance (i.e., $s_{CDS}^{(k)}, s_{SDS}^{(k)} \in \{1, 2, 3\}$) over a 50-year effective exposure horizon ($s_{\tau}^{(k)} \in \{1, 2, \dots, 50\}$). At each decision epoch, the action space for each component consists of three alternatives. The `Do Nothing' action incurs no immediate cost, permitting the component to evolve according to natural deterioration models. A `Minor Repair', costing 20\% of the replacement value, improves both CDS and SDS by one state (where applicable) and reduces the effective exposure time $\tau$ by 10 years, thus lowering future deterioration rates. Finally, a `Major Repair' restores the component to its pristine condition at a cost equal to 100\% of its replacement value. To account for economies of scale, such as shared mobilization and site preparation efficiencies, simultaneous interventions across the system are discounted by 5\% for two concurrent actions and 10\% for three. A typical discount factor of $\gamma = 0.95$ is used in this study to reflect the time value of money over the planning horizon.

The size of the state space for a single component is $|\mathcal{S}^{(k)}| = n_{CDS} \times n_{SDS} \times n_{\tau} = 3 \times 3 \times 50 = 450$ states. The total system state is the Cartesian product of the individual component states, $s = (s^{(A)}, s^{(B)}, s^{(C)})$, and the system state space is $\mathcal{S} = \mathcal{S}^{(A)} \times \mathcal{S}^{(B)} \times \mathcal{S}^{(C)}$. This leads to a total number of system states of $|\mathcal{S}| = 450^3 = 91,125,000$. The size of this state space exemplifies the curse of dimensionality inherent in nonstationary system-level infrastructure management and motivates the need for the efficient solution methodology proposed in this paper.

With three seismic and three corrosion states per component, the system seismic and corrosion configurations number $n_{\mathbf{SDS}} = n_{\mathbf{CDS}} = 3^3 = 27$, so that the joint seismic transition $\mathbf{P}_{SDS}$ is described by only $19{,}683$ entries, against a full system transition matrix of nominal size $91{,}125{,}000 \times 91{,}125{,}000$. Within a single backup, the joint seismic update contributes $n_{\mathbf{SDS}} = 27$ operations per state, compared to $\sum_{k} |\mathcal{S}^{(k)}| = 1{,}350$ for the separable factors, so that the explicit treatment of spatial dependence adds only a small fraction to the cost of a backup.

\section{Results and Discussion}
\noindent
This section presents the results obtained from applying the proposed framework to the presented infrastructure network in San Francisco. We first visualize the developed state-dependent fragility curves, which explicitly embed the time-varying vulnerability of the structures due to aging and deterioration. We then discuss the computational performance of our tensor-based value iteration method compared to a standard MDP solver. Subsequently, we analyze the effectiveness of the derived optimal maintenance policy through simulated life-cycle trajectories and a comparative cost analysis against alternative maintenance strategies.

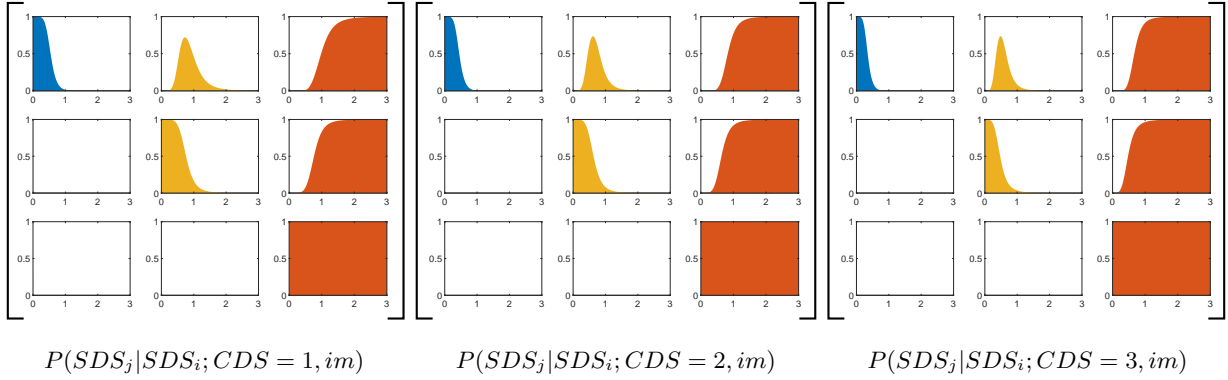
\begin{figure*}[tb]
    \centering
    \begin{tikzpicture}
        \def\mysep{0.25cm}       
        \def\outerarm{6pt}       
        \def\innerarm{6pt}    
        \def\bracketsep{2.5pt}  
        \def\labelsep{8pt}      

        \node (p1) {\includegraphics[width=0.3\textwidth]{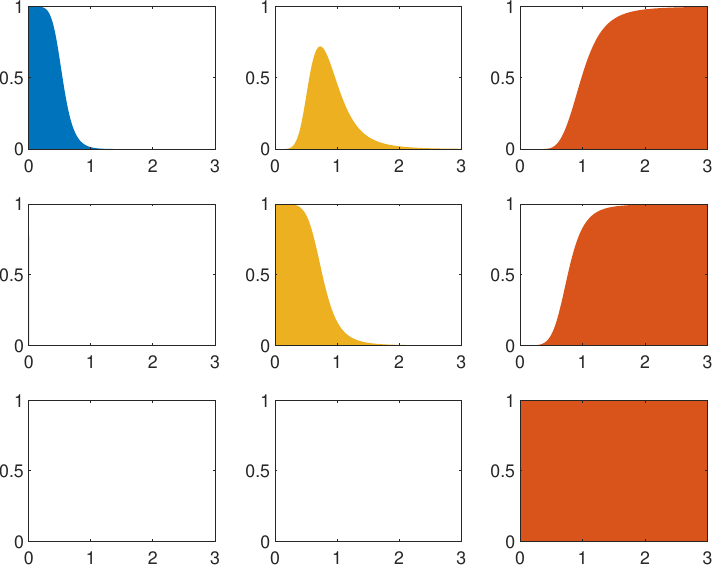}};
        \node[anchor=west] (p2) at ([xshift=\mysep]p1.east) {\includegraphics[width=0.3\textwidth]{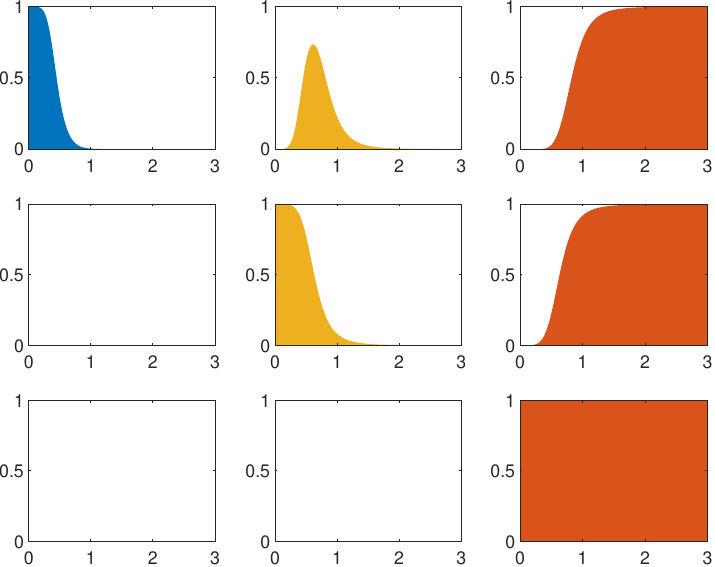}};
        \node[anchor=west] (p3) at ([xshift=\mysep]p2.east) {\includegraphics[width=0.3\textwidth]{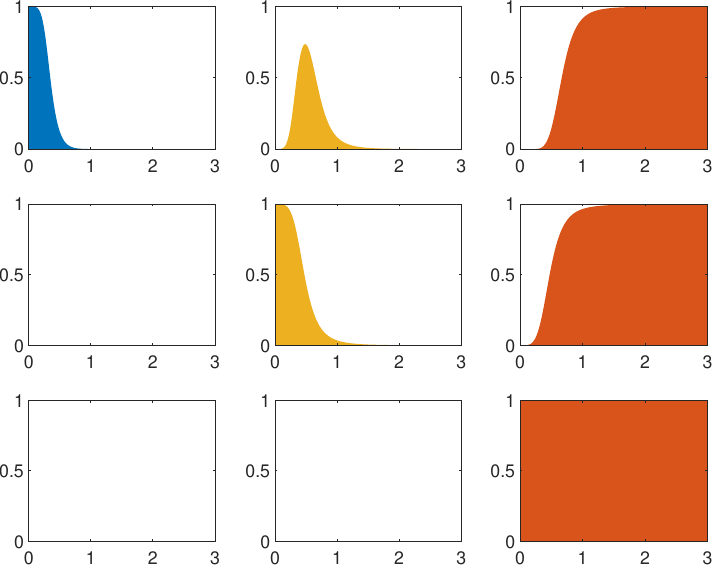}};

        \node[below=\labelsep of p1] {\footnotesize $P(SDS_j | SDS_i; CDS = 1, im)$};
        \node[below=\labelsep of p2] {\footnotesize $P(SDS_j | SDS_i; CDS = 2, im)$};
        \node[below=\labelsep of p3] {\footnotesize $P(SDS_j | SDS_i; CDS = 3, im)$};

        \draw[thick] 
            ($ (p1.north west) + (\outerarm,0) $) -- (p1.north west) -- (p1.south west) -- ($ (p1.south west) + (\outerarm,0) $);
        \draw[thick] 
            ($ (p3.north east) + (-\outerarm,0) $) -- (p3.north east) -- (p3.south east) -- ($ (p3.south east) + (-\outerarm,0) $);
            
        \coordinate (topmid1) at ($ (p1.north east)!0.5!(p2.north west) $);
        \coordinate (botmid1) at ($ (p1.south east)!0.5!(p2.south west) $);

        \draw[thick] ($ (topmid1) - (\bracketsep,0) + (-\innerarm,0) $) -- ($ (topmid1) - (\bracketsep,0) $) -- ($ (botmid1) - (\bracketsep,0) $) -- ($ (botmid1) - (\bracketsep,0) + (-\innerarm,0) $);
        \draw[thick] ($ (topmid1) + (\bracketsep,0) + (\innerarm,0) $) -- ($ (topmid1) + (\bracketsep,0) $) -- ($ (botmid1) + (\bracketsep,0) $) -- ($ (botmid1) + (\bracketsep,0) + (\innerarm,0) $);

        \coordinate (topmid2) at ($ (p2.north east)!0.5!(p3.north west) $);
        \coordinate (botmid2) at ($ (p2.south east)!0.5!(p3.south west) $);

        \draw[thick] ($ (topmid2) - (\bracketsep,0) + (-\innerarm,0) $) -- ($ (topmid2) - (\bracketsep,0) $) -- ($ (botmid2) - (\bracketsep,0) $) -- ($ (botmid2) - (\bracketsep,0) + (-\innerarm,0) $);
        \draw[thick] ($ (topmid2) + (\bracketsep,0) + (\innerarm,0) $) -- ($ (topmid2) + (\bracketsep,0) $) -- ($ (botmid2) + (\bracketsep,0) $) -- ($ (botmid2) + (\bracketsep,0) + (\innerarm,0) $);

    \end{tikzpicture}
    
    \caption{IM-conditional state transition probability matrix showing the probability of transitioning to a seismic damage state $SDS_j$ (columns) from an initial state $SDS_i$ (rows), conditioned on the corrosion damage state ($CDS=1, 2, 3$) and $IM$. Each matrix of plots corresponds to a different level of pre-existing corrosion.}
    \label{fig:fragility_matrix}
\end{figure*}
\begin{figure*}[tb]
    \centering
    \begin{subfigure}[b]{0.35\textwidth}
        \centering
        \includegraphics[width=\textwidth]{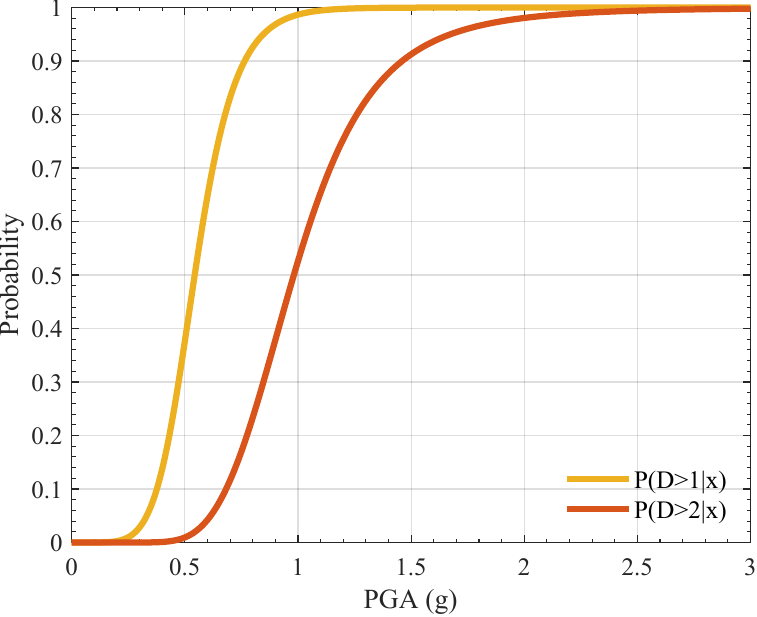} 
        \caption{\footnotesize $P(SDS_j|SDS_i=1,CDS=1,im)$}
        \label{fig:frag_sds1}
    \end{subfigure}
    \begin{subfigure}[b]{0.35\textwidth}
        \centering
        \includegraphics[width=\textwidth]{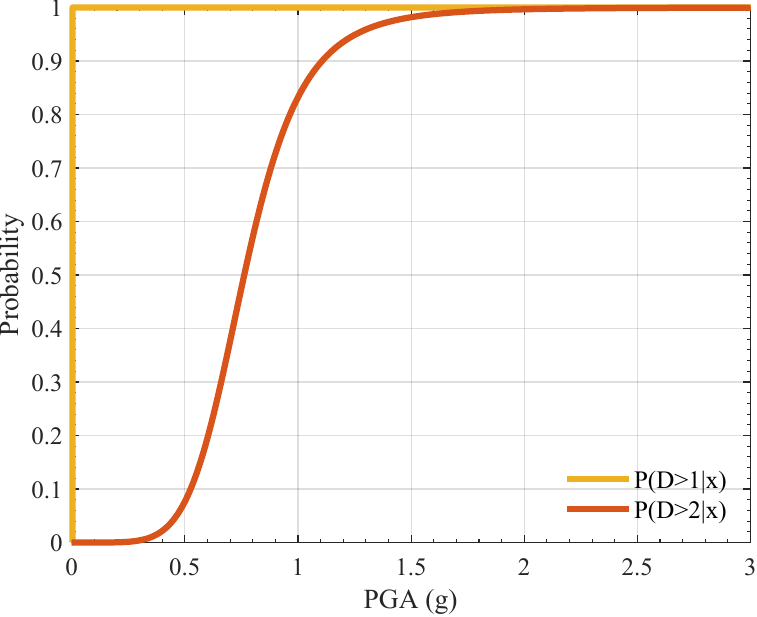} 
        \caption{\footnotesize $P(SDS_j|SDS_i=2, CDS=1, im)$}
        \label{fig:frag_sds2}
    \end{subfigure}
   \caption{State-dependent fragility plots for a component with initial corrosion damage state $CDS=1$. Each plot shows the probability of transitioning to a final seismic damage state $SDS_j$ conditioned on the initial state $SDS_i$.}
    \label{fig:fragility_plots}
\end{figure*}

\subsection{State-Dependent Fragility}
\noindent
The probabilistic link between seismic hazard intensity and structural damage is captured by the state-dependent models, $P(SDS_t = j | SDS_{t-1} = i, CDS_t = l, IM_t)$. These models graphically represent the IM-conditional transition probabilities, linking the structure’s physical condition to its seismic vulnerability. To populate these crucial transition probabilities, a comprehensive simulation campaign was conducted following the workflow described in Section~\ref{sec:workflow}. A total of 5,000 Monte Carlo simulations of the 50-year life-cycle trajectory were performed for each building archetype. In each realization, seismic events have been generated in time as a Poisson process based on the site's mean annual rate, and, for each event, the intensity measure has been sampled from the site-specific hazard curve, with a compatible ground motion then selected~\citep{vlachos2016multi, vlachos2018predictive} and applied in nonlinear time-history analysis, following the workflow of Section~\ref{sec:workflow}. This number of simulations ensures statistical robustness and provides a rich dataset covering a wide range of deterioration levels and seismic intensities. The resulting data on structural response ($D_{PA}$) have been used to fit the parameters of the fragility models via the maximum likelihood estimation approach detailed in Section~\ref{sec:statefrag}.

The resulting family of IM-conditional state transition probability matrix, as illustrated in Figure~\ref{fig:fragility_matrix}, defines the SDS transition probabilities conditioned on both the IM and the pre-existing CDS. The figure is organized into three primary blocks, each corresponding to a specific level of corrosion. Within each block, the rows denote the initial seismic damage state ($SDS_i$), while the columns represent the final state ($SDS_j$). Each individual plot shows the transition probability (y-axis) as a function of the IM (x-axis). For instance, the first row of any matrix details the probabilities for a component initially in $SDS_1$, the first plot shows the probability of remaining in $SDS_1$, while the second and third plots show the probabilities of transitioning to $SDS_2$ and $SDS_3$, respectively, for a given IM.

A key observation from Figure~\ref{fig:fragility_matrix} is the pronounced effect of pre-existing corrosion on the seismic vulnerability of the component. As the CDS progresses from state 1 to 3, the probability density functions for damage state transitions exhibit a distinct leftward shift. This shift signifies that for any given IM, a more heavily deteriorated component is significantly more vulnerable to sustaining severe
damage. Consequently, the likelihood of remaining in a lower damage state (e.g., $SDS_1$ or $SDS_2$) diminishes as the corrosion degradation increases, highlighting the critical coupling between gradual deterioration and acute seismic vulnerability.

These IM-conditional transition probability matrices can then be used to derive the state-dependent fragility functions. Figure~\ref{fig:fragility_plots} illustrates these fragility plots for the case where the component has an initial corrosion damage state of $CDS=1$. The two plots correspond to the probabilities of the final seismic damage state ($SDS_j$) given an initial state of $SDS_i=1$ and $2$. The plots confirm that seismic damage is non-decreasing within an event, consistent with the $j \ge i$ constraint imposed in Equation~\ref{eq:log_likelihood_latex}. Consequently, transitions from a higher damage state to a lower one (e.g., from $SDS_3$ to $SDS_2$) are not possible, as repair actions are not considered in this assessment.

\begin{figure*}[!b]
    \centering
    \begin{subfigure}{0.48\textwidth}
        \centering
        \includegraphics[width=\textwidth]{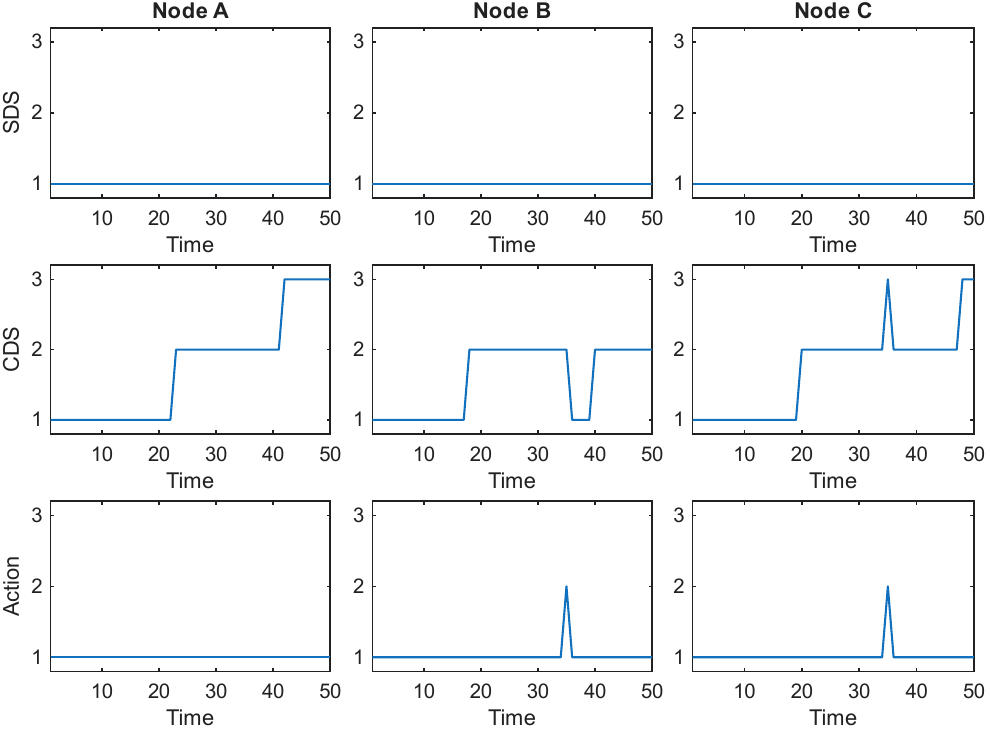}
        \caption{} 
        \label{fig:real1}
    \end{subfigure}
        \hfill
    \begin{subfigure}{0.48\textwidth}
        \centering
        \includegraphics[width=\textwidth]{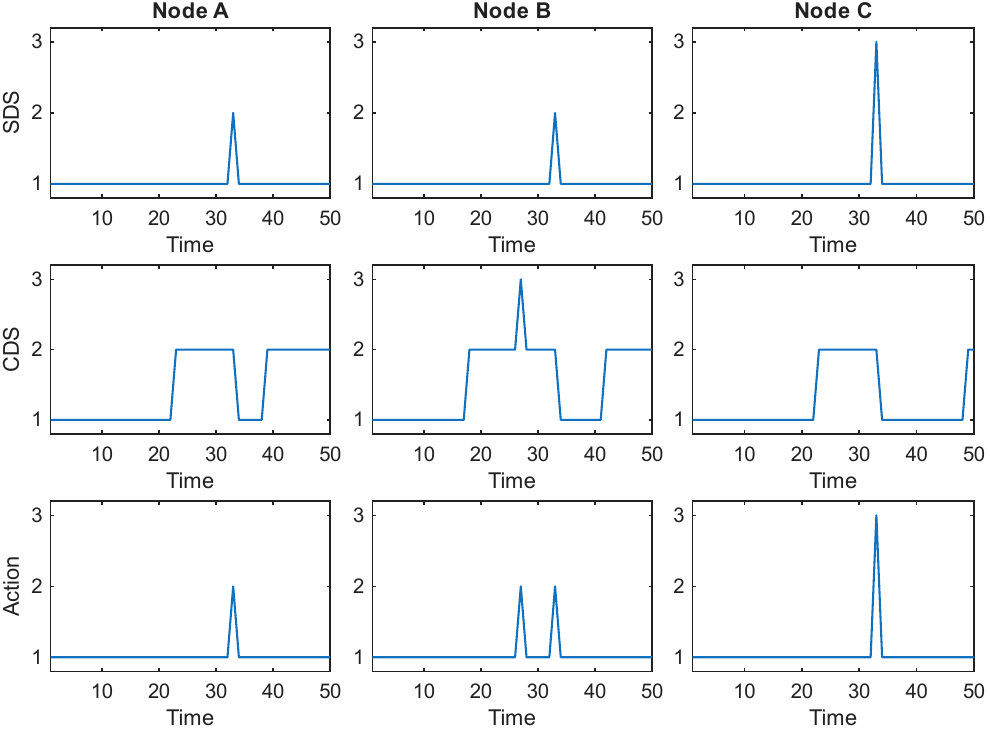}
        \caption{} 
        \label{fig:real2}
    \end{subfigure}

    \caption{Sample life-cycle realizations showing maintenance actions, CDS, and SDS over 50 years under the optimal MDP policy.}
    \label{fig:sample_realizations}
\end{figure*}

\subsection{Computational Performance}
\noindent The finite-horizon MDP discussed above is solved using two distinct approaches: a
standard MDP solver (specifically, the \texttt{MDPToolbox} for MATLAB
\citep{chades2014mdptoolbox}) and our novel tensor-based method. \textcolor{black}{The
two solvers are first compared on the separable formulation, in which the components are
treated as independent and both solvers are applicable.} On a local workstation
equipped with an Intel\textsuperscript{\textregistered}
Xeon\textsuperscript{\textregistered} Platinum 8260 CPU @ 2.40GHz, our method \textcolor{black}{obtained the exact
optimal policy in approximately 45 minutes, whereas the \texttt{MDPToolbox}, despite
leveraging sparse matrix representations, required approximately 500 GB of memory and 14 hours to converge
to the identical solution.}

\textcolor{black}{The advantage becomes even more decisive for the correlated formulation. The
joint seismic transition of Equation~\ref{eq:joint_seismic} substantially increases
the number of non-zero entries per row of $\mathbf{P}_a$, and the sparse system matrices
required by the \texttt{MDPToolbox} exceeded our available 700 GB of memory, so that the
problem could not be solved. Our proposed method, by contrast,
never forms these matrices and solves the same problem exactly within a peak memory
footprint of just 8 GB. This is a direct consequence of the structure exploitation in
Section~\ref{sec:coupled}: the coupling is stored in the $19{,}683$ entries of
$\mathbf{P}_{SDS}$ and applied on the seismic damage subspace, while the remaining dynamics are
handled through mode-wise products on the component matrices. Spatial correlation is
thus fully incorporated without any loss of accuracy and at a memory cost that remains
negligible relative to the nominal size of the problem. This substantial reduction in both computation time and memory validates the practical utility of our approach, making it feasible to apply the framework to larger-scale, multi-component infrastructure systems that could otherwise have intractable solutions.}

\subsection{Performance of the Optimal Policy}
\noindent
The performance of the optimal policy is evaluated through life-cycle simulations, focusing on its efficacy in both risk reduction and economic efficiency. All results in this section correspond to the partially coupled formulation of Section~\ref{sec:coupled}, solved with the proposed tensor-based method. This simulation-based demonstrations provide insights into how the policy adapts to evolving deterioration and hazard risks. 

\subsubsection{Sample Life-Cycle Realizations}
\noindent
Before presenting aggregate performance metrics, we first illustrate how the optimal policy behaves on two representative life-cycle trajectories. Figure~\ref{fig:sample_realizations} depicts two sample realizations of the system's life-cycle evolution under the globally optimal policy, showing the chosen actions, CDS, and SDS for Nodes A, B, and C over the 50-year design life. These realizations offer valuable insights into the adaptive nature of the optimal policy.

\begin{figure}[!b]
    \centering
    \includegraphics[width=0.45\linewidth]{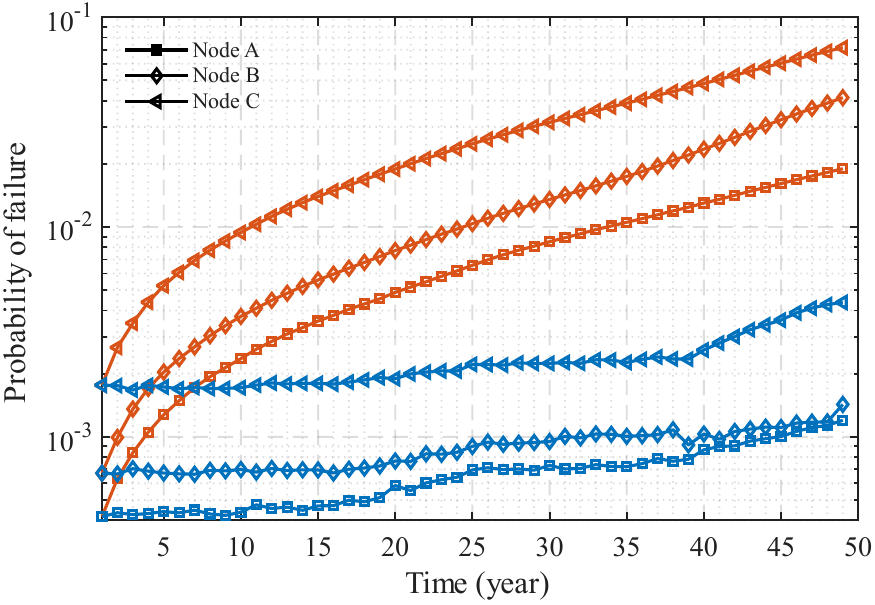}
    \caption{Variation of probability of failure over time for all nodes. The red curves represent the case with no maintenance actions, while the black curves show the outcomes under the MDP optimal policy.}
    \label{fig:failure_prob_nodeA}
\end{figure}

In the first realization, in Figure~\ref{fig:sample_realizations}a, no damaging seismic event occurs and the system evolves under corrosion deterioration alone. All three nodes advance to CDS 2 between years 17 and 22 without prompting intervention, since deterioration carries no direct cost and acts only by indirectly elevating the state-dependent seismic vulnerability. The policy intervenes only when Node C reaches CDS 3 in year 34, and it then repairs Nodes C and B together. Treating Node B, still at CDS 2, is not merely opportunistic. Node C supports both other nodes in the dependency structure of Figure~\ref{fig:sf_system}b, and this dependence enters the optimization through the super-additive joint failure costs of Table~\ref{tab:failure_costs}, so that the deteriorated condition of C raises the value of improving B as well. The concurrent intervention also attracts a campaign discount and reduces Node B's effective exposure time. Node A is left untreated and reaches CDS 3 in year 42.

The second realization, in Figure~\ref{fig:sample_realizations}b, illustrates the response to a seismic hazard event. Node B first reaches CDS 3 in year 28 and receives an isolated Minor Repair. In year 33, a single regional event then damages all three nodes at once, driving Nodes A and B to SDS 2 and Node C to its failure state, SDS 3. The policy responds in the same epoch with a coordinated campaign across all three nodes. Node C receives a Major Repair, while Nodes A and B are returned to SDS 1 with Minor Repair, with the three concurrent interventions securing the largest campaign discount. Toward the end of the horizon, corrosion again advances all three nodes to CDS 2 without triggering further action, since the risk reduction achievable over the short remaining service life no longer justifies the cost of intervention. This behavior reflects the zero-value terminal state adopted here. The framework can readily also accommodate other temporal boundary conditions, such as residual asset value or state-dependent decommissioning costs, which would affect decisions near the end of the planning horizon. Additional representative realizations are also provided in \ref{app:realizations}.

Having illustrated the mechanism qualitatively, we now quantify performance across Monte Carlo simulations using safety, risk, and cost metrics. First, the policy's ability to enhance system safety is demonstrated by comparing the component failure probability against a baseline 'do-nothing' scenario. As shown in Figure~\ref{fig:failure_prob_nodeA}, the optimal policy significantly reduces the cumulative probability of failure over the 50-year service life across all three nodes. This demonstrates the effectiveness of the framework in proactively managing deterioration and mitigating seismic risks.

\begin{figure}[!tb]
    \centering
    \includegraphics[width=0.45\linewidth]{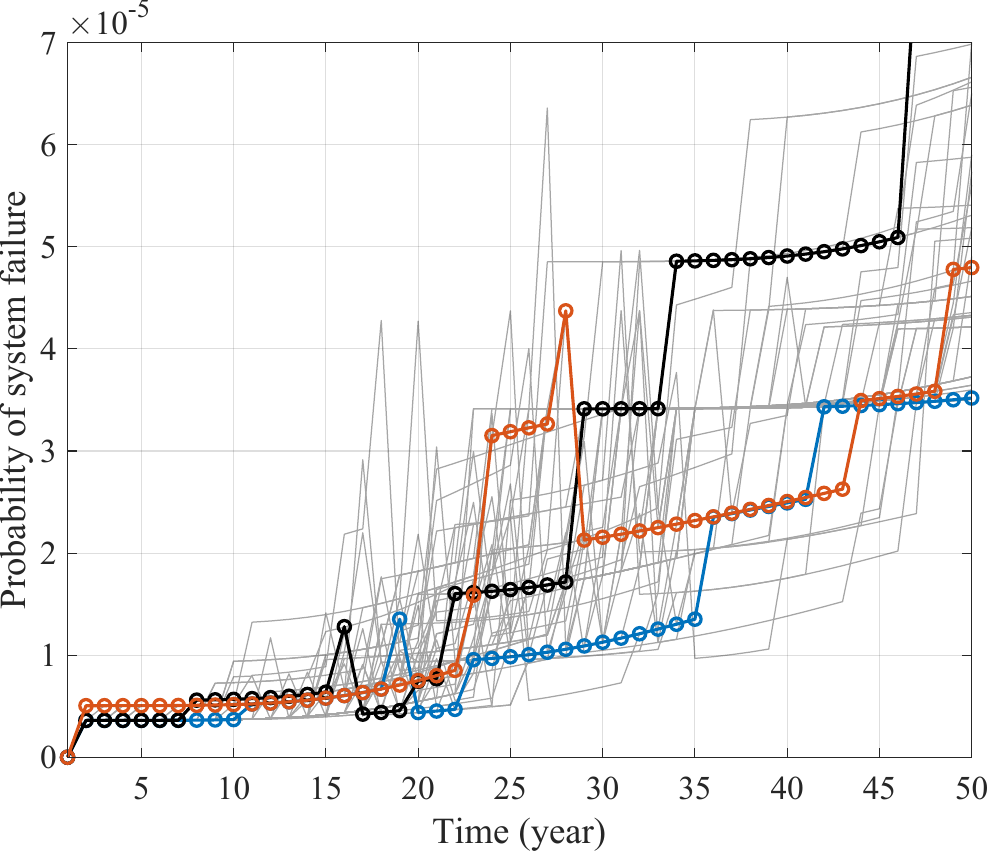}
    \caption{Sample trajectories of system-level risk evolution over 50 years under the optimal policy.}
    \label{fig:system_risk}
\end{figure}

To reconnect with the conceptual framework introduced in Figure~\ref{fig:performance_diagrams}, we visualize the total system risk, defined here as the joint failure probability of Nodes A, B, and C, over the entire 50-year horizon under the optimal MDP policy. As shown in Figure~\ref{fig:system_risk}, risk evolves dynamically in response to both progressive deterioration and shock-based events, closely mirroring the hypothesized risk trajectories in the introductory diagram. Figure~\ref{fig:system_risk} displays numerous possible risk evolutions as grayed-out lines, with three representative trajectories highlighted to illustrate various potential outcomes. The variation among these trajectories arises because the optimal policy is state-dependent and the system evolution is stochastic. The decision to intervene is not based on a simple, fixed risk threshold; instead, the MDP policy evaluates the complete system state, factoring in the condition of each individual component, their interdependencies, and the remaining time in the system’s life-cycle. The results clearly demonstrate how proactive, adaptive interventions mitigate system-wide risk, thereby enhancing resilience throughout the life-cycle.

\begin{figure}[!tb]
    \centering
    \includegraphics[width=0.45\linewidth]{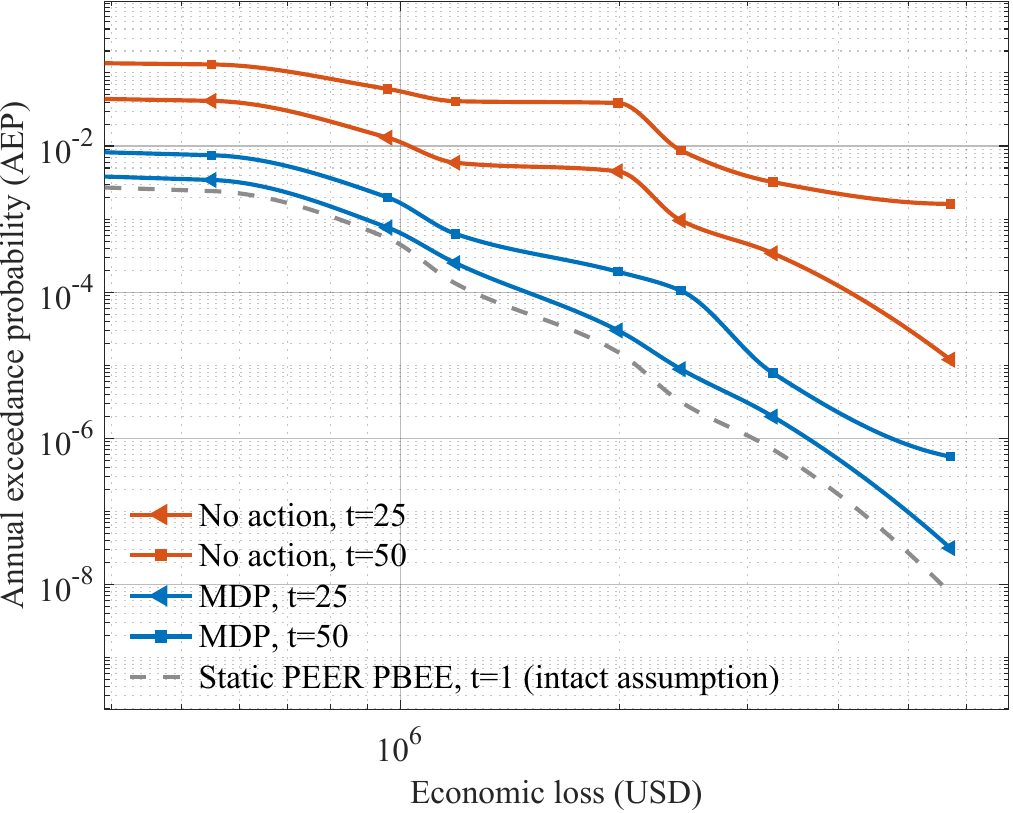}
    \caption{Annual exceedance probability (AEP) vs. economic loss for different management policies. The static PEER PBEE baseline (dashed gray) assumes a pristine, time-invariant system and is the same as the No action policy at t=1. As the system ages (solid curves), the risk under the No action policy (orange) increases from t=25 to t=50. In contrast, the adaptive MDP policy (black) consistently maintains a lower AEP at both time points, showing effective risk reduction through maintenance.}
    \label{fig:aep_loss}
\end{figure}

\begin{table*}[b]
\centering
\caption{Definitions of the Condition-Based Maintenance (CBM) baseline policies and their intervention triggers.}
\label{tab:cbm_rules}
\begin{tabular}{@{} l l l l @{}}
\toprule
\textbf{Policy} & \textbf{Do Nothing} & \textbf{Minor Repair} & \textbf{Major Repair} \\ 
\midrule
CBM Policy \#1 & $CDS \le 2$ and $SDS = 1$ & $SDS = 2$ & $CDS = 3$ or $SDS = 3$ \\
CBM Policy \#2 & $CDS = 1$ and $SDS = 1$ & $CDS = 2$ & $CDS = 3$ or $SDS \ge 2$ \\
CBM Policy \#3 & $CDS \le 2$ and $SDS \le 2$ & $CDS = 3$ &  $SDS = 3$ \\ 
\bottomrule
\end{tabular}
\end{table*}

We also compare the standard PEER PBEE output, the plot of annual exceedance probability (AEP) versus economic loss, with our time-aware framework. Since our model evolves with age, we plot AEP–loss curves at two snapshots ($t{=}25$ and $t{=}50$ years) and contrast the optimal MDP policy against the static PEER PBEE outcome and a no action scenario. As shown in Figure~\ref{fig:aep_loss}, the static PEER PBEE curve lies below the time-aware curves as it assumes a pristine, time-invariant condition, thus underestimating the risk. The No-action curves shift upward from $t{=}25$ to $t{=}50$, indicating that aging increases the probability of exceeding any loss threshold. In contrast, the MDP curves stay consistently lower, reducing exceedance across the entire loss range and reinforcing a more economical, adaptive, and realistic maintenance policy.

\begin{figure}[!tb]
    \centering
    \includegraphics[width=0.45\linewidth]{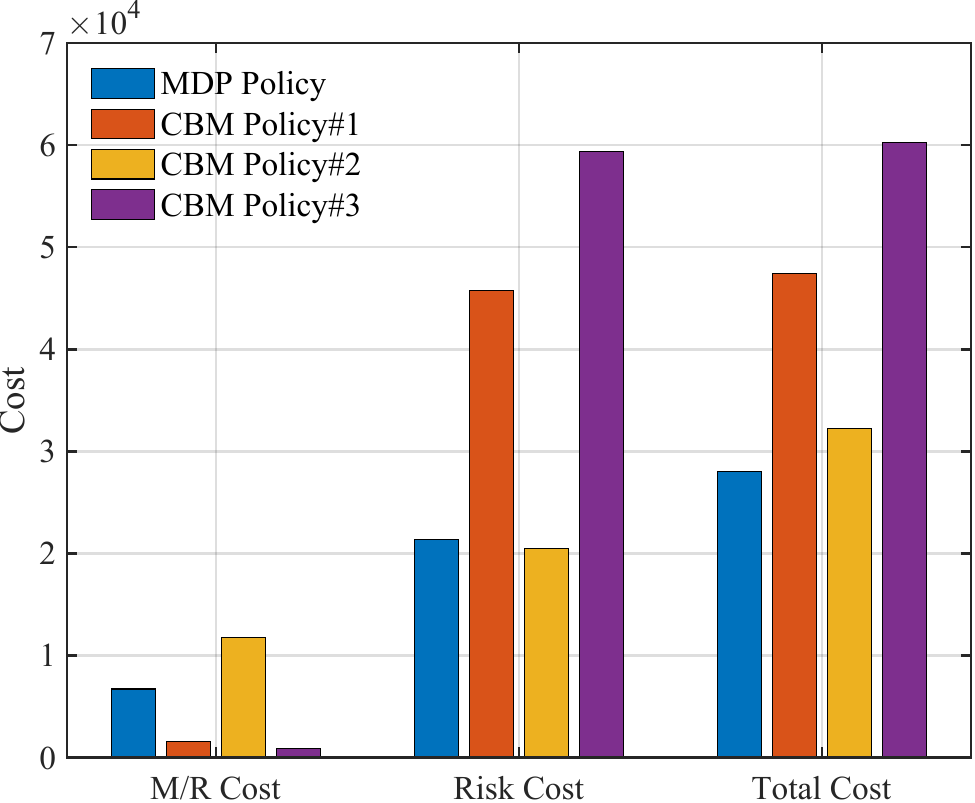}
    \caption{Comparison of total life-cycle costs, including maintenance/repair (M/R) costs and risk costs, for the optimal MDP policy and three different top-performing condition-based maintenance (CBM) policies.}
    \label{fig:cost_comparison}
\end{figure}

To further validate the optimality of the derived policy, we compare the total life-cycle cost against three representative condition-based maintenance (CBM) policies. These CBM policies represent carefully selected, top-performing strategies from a broad candidate pool, where maintenance is triggered based on observed damage states. Table~\ref{tab:cbm_rules} summarizes the intervention rules defining each policy. Figure~\ref{fig:cost_comparison} illustrates this comparison, showing the breakdown of maintenance/repair (M/R) costs, risk costs (expected failure costs), and the total life-cycle cost. As evident from the figure, the total cost incurred when following the globally optimal MDP policy (represented by the black bar) is consistently the lowest among all compared policies.

The CBM policy \#2 that involves frequent repair and maintenance (represented by the yellow bar) incurs significantly higher M/R costs. While this strategy leads to somewhat lower risk costs due to improved component conditions, the overall total cost is still higher than that achieved by the MDP. This indicates that overly aggressive maintenance, without a holistic view of future risks and costs, can be economically inefficient. Similarly, the CBM policy \#1 (represented by the red bar) attempts a compromise by tolerating moderate corrosion. While it better balance M/R and risk costs, its rigid static thresholds fail to adapt to the system's aging trajectory, ultimately costing more than the dynamically optimized MDP. Conversely, the CBM policy \#3 that involves less frequent repair and maintenance actions (represented by the purple bar) results in lower M/R costs. However, this comes at the expense of very high risk costs due to a higher probability of component and system failures. Consequently, the total cost for this strategy is also higher than the MDP's. This highlights the dangers of insufficient maintenance in the face of deteriorating conditions and hazard risks.

In summary, this section demonstrates the practical side of the proposed framework. Since the MDP policy is globally optimal by construction, no other policy can achieve a lower expected life-cycle cost, and the CBM comparisons serve to also illustrate this guarantee, as the indicative results confirm.

\section{Conclusions}
\noindent
This paper advances the established probabilistic hazard analysis paradigm from a static risk quantification tool into a dynamic, decision-theoretic approach. The conventional PEER PBEE framework evaluates structural performance assuming a fixed, typically pristine condition state, lacking a formal mechanism to integrate time-dependent degradation or structural interventions. Meanwhile, existing maintenance and life-cycle frameworks explicitly address the time dimension but fail to embed hazards in a fully probabilistic, risk-based form consistent with established methodologies, such as Probabilistic Seismic Hazard Analysis (PSHA). The proposed framework bridges this disconnect by jointly capturing the continuous evolution of risk, driven by nonstationary deterioration and cumulative hazard damage, and embedding a stochastic optimal control mechanism to systematically manage that evolving risk over the asset's lifespan. To realize this, the framework models continuous degradation via nonstationary processes and integrates it with state-dependent generalized fragility functions within a Dynamic Bayesian Network (DBN), capturing the coupled temporal evolution of structural vulnerability and hazard exposure. This time-aware probabilistic assessment is then formally cast as a finite-horizon Markov Decision Process (MDP), yielding globally optimal, adaptive maintenance policies that minimize total expected life-cycle costs with mathematical guarantees.

The practical application and effectiveness of the proposed framework is demonstrated through a representative numerical example involving a critical infrastructure network in San Francisco, California. By assuming a time-invariant, pristine condition, the static PEER PBEE baseline underestimates economic risk, a gap that widens as the system ages. Furthermore, because the MDP formulation mathematically guarantees a globally optimal policy, it inherently outperforms traditional condition-based maintenance (CBM) policies. The comparative cost analysis serves to illustrate this theoretical superiority, exposing the shortcomings of methods unable to adapt to the jointly evolving states of deterioration, shock damage, and system dependencies. In contrast, the proposed framework can provide adaptive interventions informed by the probabilistic hazard environment and the system’s evolving conditions. Naturally, the presented example is only intended to demonstrate the framework on a general and realistic setting, rather than to exhaust the modeling details available within it. The structural and deterioration models, the hazard characterization and its fragility functions, and the consequence model, all supply inputs to the formulation and each can be refined or specialized to the case at hand, without altering the overall newly developed framework, the dynamic risk formulation, the DBN, the MDP, or the tensor-based solver components.

A known critical challenge in system-level infrastructure management, the curse of dimensionality, is mitigated here by developing a novel tensor-based computational method that exploits Kronecker-factored transition dynamics and tensor algebra to efficiently solve the high-dimensional MDP. The same construction extends to partially coupled dynamics induced by spatially correlated hazards, for which the coupling is confined to a low-dimensional subspace and the exactness of the solution is preserved. This approach yields substantial reductions in computational time and memory requirements relative to conventional sparse solvers, while still preserving the accuracy of exact dynamic programming.

Overall, this framework provides asset managers and engineers with a quantitative, practical decision-support tool, advancing the PEER PBEE methodology by integrating it with strategic, time-aware, life-cycle planning toward cost-effective risk mitigation and safer infrastructure systems. It offers a comprehensive study of infrastructure performance over time, accounting for aging, stochastic hazards, and optimal interventions within a single, cohesive decision-making framework.

{\footnotesize
\bibliographystyle{elsarticle-num} 
\bibliography{ref}
}
\clearpage
\appendix
\setcounter{figure}{0}
\renewcommand{\thefigure}{A.\arabic{figure}}

\begin{figure*}[!b]     
  \centering
  \subcaptionbox{\label{fig:real3}}{\includegraphics[width=0.45\textwidth]{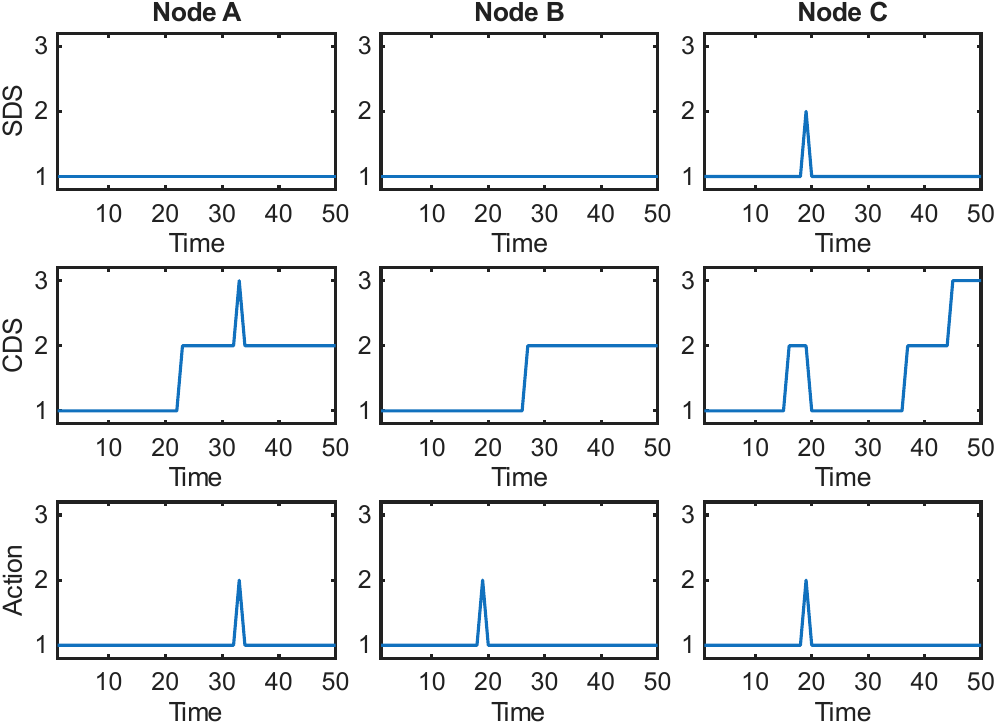}}\hfill
  \subcaptionbox{\label{fig:real4}}{\includegraphics[width=0.45\textwidth]{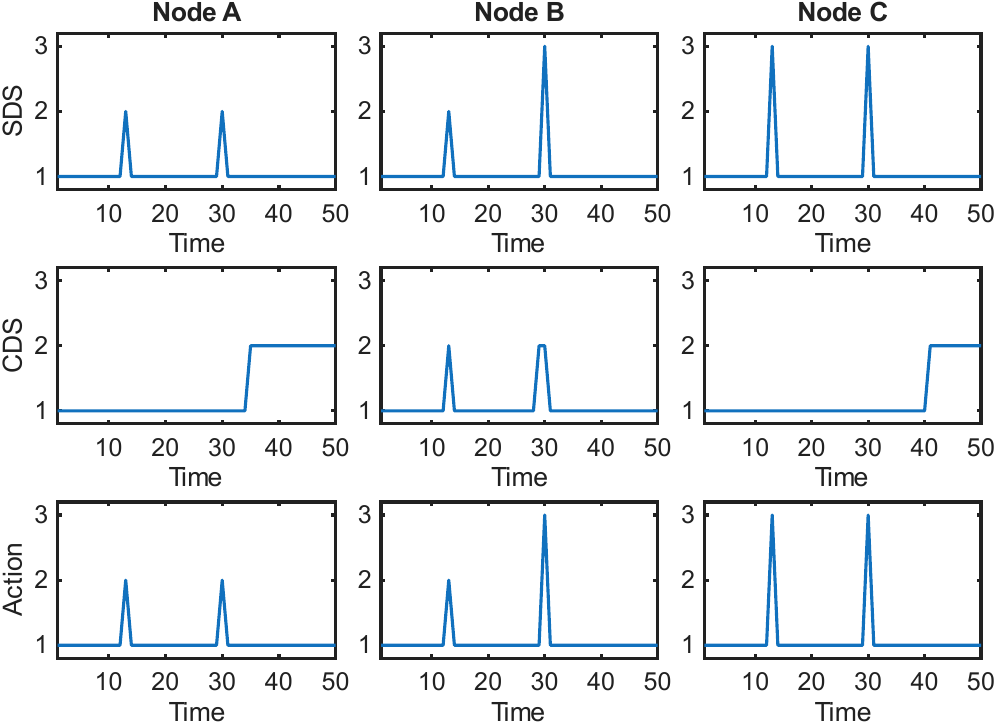}}
    \subcaptionbox{\label{fig:real5}}{\includegraphics[width=0.45\textwidth]{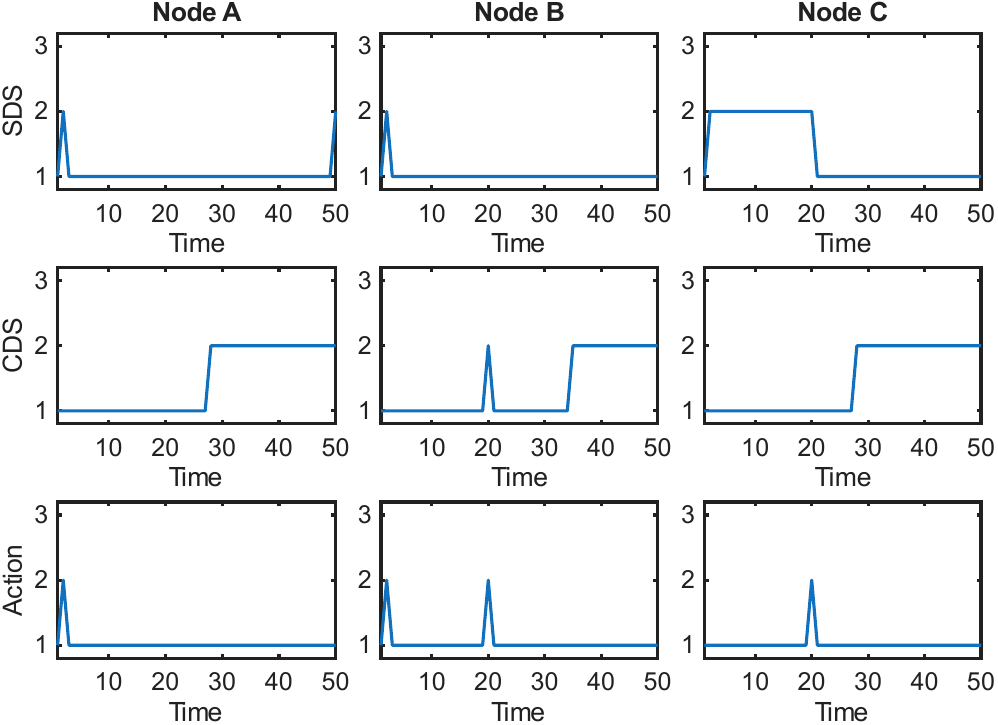}}\hfill
  \subcaptionbox{\label{fig:real6}}{\includegraphics[width=0.45\textwidth]{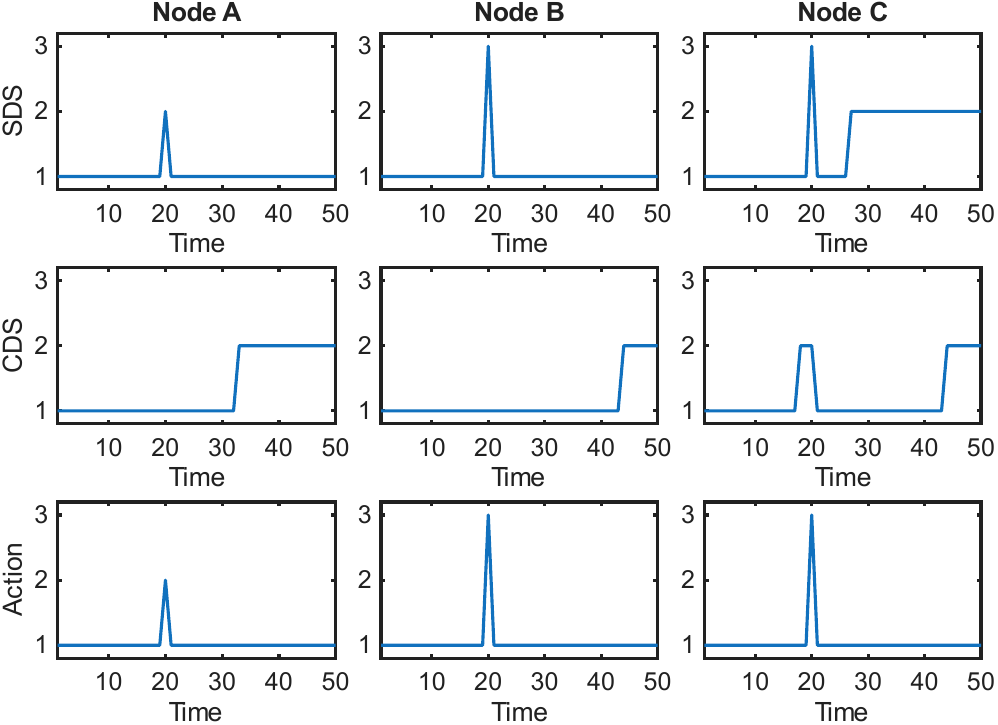}}
  \caption{Additional sample life-cycle realizations under the optimal MDP policy, showing maintenance actions, CDS, and SDS for Nodes A, B, and C over the 50-year horizon.}
  \label{fig:appendix_realizations}
\end{figure*}

\section{Additional Life-Cycle Realizations}
\label{app:realizations}
\noindent
To complement the two trajectories presented in the main text (Figure~\ref{fig:sample_realizations}), this appendix reports four additional representative realizations that highlight distinct facets of the optimal policy and of the coupled deterioration--hazard dynamics. Figure~\ref{fig:real3} isolates the effect of corrosion-dependent fragility. A single seismic event near year~19 drives only Node~C to $SDS=2$, while Nodes~A and~B remain at $SDS=1$. Node~C was already at $CDS=2$, against $CDS=1$ for the others, so its elevated corrosion raised its seismic vulnerability through the state-dependent fragility. The policy applies a `Minor Repair', reducing its seismic and corrosion damage states.

Figure~\ref{fig:real4} showcases a response to repeated shocks. Two events, near years~13 and~30, inflict heterogeneous damage. The first event drives Nodes~A and~B to $SDS=2$ and Node~C to failure ($SDS=3$), while the second drives Nodes~B and~C to failure and Node~A to $SDS=2$. In each case the policy applies `Minor Repair' to the moderately damaged components and `Major Repair' to the failed ones, restoring serviceability. This example concisely illustrates coordinated, component-specific intervention under multi-event, dependent exposure.

Figure~\ref{fig:real5} illustrates deferred, co-scheduled maintenance. A regional event at year~1 drives all three nodes to $SDS=2$. The policy restores Nodes~A and~B with `Minor Repair' but defers Node~C repairs, keeping it at this moderate, non-failure state until year~20, when Node~B's corrosion state reaches $CDS=2$ and then the two are repaired together, securing also the campaign discount.

Finally, Figure~\ref{fig:real6} displays various damage response strategies. A seismic event at year~20 fails Nodes~B and~C and drives Node~A to $SDS=2$, with all nodes receiving repairs and returning to $SDS=1$. A later seismic event then only affects Node~C this time and drives it to $SDS=2$. However, this moderate damage is left unrepaired now for the remainder of the service life.

\end{document}